\documentclass[11pt]{article}

\usepackage[preprint]{acl}

\usepackage{times}
\usepackage{latexsym}

\usepackage[T1]{fontenc}
\usepackage[utf8]{inputenc}

\usepackage{microtype}

\usepackage{inconsolata}

\usepackage{graphicx}

\usepackage{booktabs}
\usepackage{multirow}
\usepackage{subcaption}
\usepackage{amsmath, amssymb}
\usepackage{xcolor}
\usepackage{url}
\usepackage{fvextra}
\fvset{fontsize=\small, breaklines=true, frame=single}
\usepackage[defaultcolor=red]{changes}
\usepackage[normalem]{ulem}

\DefineVerbatimEnvironment{PromptVerb}{Verbatim}{%
  fontsize=\footnotesize,%
  breaklines=true,%
  breakanywhere=true,%
  breakafter={\space},%
  breaksymbolleft={\color{gray}\tiny\ensuremath{\hookrightarrow}},%
  breakindent=0pt%
}

\title{GRASP: Plan-Guided Graph Retrieval with Adaptive Fusion and Reranking on Semi-Structured Knowledge Bases}

\author{
 \textbf{Yicheng Tao\textsuperscript{1,\dag}},
 \textbf{Yiqun Wang\textsuperscript{2,\dag}},
 \textbf{Xiangchen Song\textsuperscript{1,\dag}},
 \textbf{Xin Luo\textsuperscript{2}},
 \textbf{Kai Liu\textsuperscript{2}},
 \textbf{Yuanhao Huang\textsuperscript{2}},
 \textbf{Jie Liu\textsuperscript{1,2,\dag}}
\\
 \textsuperscript{1}Department of Electrical Engineering and Computer Science, University of Michigan
\\
 \textsuperscript{2}Department of Computational Medicine and Bioinformatics, University of Michigan
\\
 \textsuperscript{\dag}Correspondence
}

\begin{document}
\maketitle
\begin{abstract}
Retrieval over semi-structured knowledge bases (SKBs) must handle both textual descriptions and relational constraints.
Text-based retrievers capture the overall semantics of a query but struggle with multi-hop relations, whereas graph-based retrievers follow relational paths precisely but miss the broader meaning.
We propose GRASP, a hybrid retriever that learns to fuse the two.
A small LLM planner translates the query into an executable graph plan in a single call, its graph ranking is fused with a text ranking, and a fine-tuned reranker reorders the fused candidates.
Instead of training the planner to be accurate on its own, we optimize it with reinforcement learning to maximize the gain its graph ranking brings to the fused ranking.
On three STaRK benchmarks, GRASP achieves the best results on all metrics, improving the average Hit@1 by 18.6 points over the strongest baseline, while taking only 2.47 seconds per query.
% Code: \url{https://anonymous.4open.science/r/GRASP-6B75}.
\end{abstract}

\section{Introduction}
\label{sec:intro}

Real-world queries usually combine free-form textual descriptions of entities with relational constraints among them.
For example, in e-commerce product search, a shopper may look for ``a push-along tricycle from Radio Flyer that is both fun and safe for my kid'', and for precision-medicine inquiries, a medical researcher may ask ``what disease is associated with the PNPLA8 gene and presents with hypotonia as a symptom?'' \citep{stark}. 
Answering such queries requires retrieval over \emph{semi-structured knowledge bases} (\textbf{SKB}s), which are collections of textual documents (e.g., product descriptions, paper abstracts, or gene and disease descriptions) represented as attributed nodes with a head name, in a richly typed graph with relations connecting them (e.g., also buying, citing, or interacting with). An example of such an SKB is shown in Figure~\ref{fig:amazon_exp}.
Methods for SKB retrieval can be mainly categorized into text-based and graph-based ones. The former treat retrieval as query--document text matching, with relevant graph information linearized into text, while the latter traverse the graph structure to find answers, with textual information as guidance.
These two families are complementary. Text-based methods capture the general semantics of the query but are less capable of following multi-hop relations, whereas graph-based methods precisely follow the graph paths the query targets but may lose its general meaning.
How to fuse them to obtain the best of both worlds remains an open question.

Existing hybrid methods combine the two with a fixed ratio~\cite{afretriever} or an LLM router~\cite{hybgrag}, without learning how to fuse them.
Moreover, trainable plan-based retrievers learn their planners only by supervised fine-tuning (SFT) on synthesized target plans~\cite{mor, graphraft}, and many graph-based methods rely on slow multi-turn LLM interactions~\cite{avatar, afretriever}.

Motivated by this, we propose GRASP, a \emph{hybrid} retriever that trains a graph retriever to learn how to be fused with a text retriever through reinforcement learning (RL).
Its graph branch uses a small LLM planner to translate the query, with the graph nodes linked from its entity mentions, into a single executable plan, whose graph ranking is fused with the ranking of a text retriever and then reordered by a fine-tuned reranker.
The planner is not trained to be accurate on its own, but to maximize the gain its graph ranking brings to the fused ranking over the text ranking alone, so that it learns to contribute where the text branch falls short.
Warm-started on only 500 teacher-written plans, it then explores all training queries with RL, using rewards computed from gold answers instead of plan annotations.

On the three STaRK benchmarks~\cite{stark}, GRASP achieves the best results on all metrics, improving the average Hit@1 by 18.6 points over the strongest baseline, and its 1.7B planner even surpasses its 27B teacher after RL.
Since the planner is called only once and the text branch runs in parallel, GRASP answers a query in 2.47 seconds on average, about 15 times faster than the strongest baseline.

\section{Related Work}
\label{sec:related}

Existing retrievers for SKBs can be broadly divided into text-based and graph-based ones, depending on whether the graph structure is explicitly accessed at retrieval time or only consumed in linearized textual form.

\paragraph{Text-based retrievers.}
Text-based retrievers cast SKB retrieval as text matching between queries and documents with sparse or dense representations, optionally enriching them with graph information linearized into text.
Classical retrievers such as sparse BM25~\cite{bm25} and dense DPR~\cite{dpr} serve as standard baselines~\cite{stark}.
With the advent of LLMs, proprietary embedding models such as ada-002~\cite{ada002} and open-source ones such as Qwen3-Embedding~\cite{qwen3} offer strong off-the-shelf alternatives.
Beyond generic retrievers, several methods are tailored to incorporate graph information.
KAR~\cite{kar} prompts an LLM to expand the query with the attributes of the nodes it mentions and of their query-relevant neighbors before dense retrieval.
mFAR~\cite{mfar} decomposes each document into fields derived from entity attributes and relations, scores the query against each field with both lexical and dense scorers, and aggregates the scores with learned query-conditioned field weights.

\paragraph{Graph-based retrievers.}
Graph-based retrievers explicitly operate on the graph structure, typically using LLMs to guide the graph exploration and conventional text matching to assess the textual relevance of nodes.
QA-GNN~\cite{qagnn} jointly reasons over the query and a local subgraph with a graph neural network.
Think-on-Graph (ToG)~\cite{tog} uses an LLM agent to iteratively explore nodes and relations on the graph via beam search.
AvaTaR~\cite{avatar} optimizes the prompts of a tool-using LLM agent with a comparator that contrasts positive and negative training queries, so that the agent better interleaves relational lookups with textual scoring.
HybGRAG~\cite{hybgrag} uses an LLM router to extract topic entities and relations and to choose between a text retrieval module and a hybrid module, which ranks the relation-constrained neighbors of topic entities by textual similarity. It then iteratively refines the routing with feedback from an LLM critic.
More recent methods first generate an explicit retrieval plan and then execute it on the graph.
MoR~\cite{mor} generates planning graphs consisting of metapaths whose entities carry textual restrictions, executes them by interleaving type-constrained traversal with textual matching, and reranks the candidates based on features of their traversal trajectories.
Its planner is trained via supervised fine-tuning~\cite[SFT]{ouyang2022training} on LLM-synthesized planning graphs, while its reranker is trained separately on all training trajectories.
GraphRAFT~\cite{graphraft} fine-tunes an LLM via SFT to generate Cypher~\cite{cypher} queries, using as targets the best-performing queries enumerated from multi-hop templates around the extracted entities, and applies graph-grounded constrained decoding at inference time to guarantee executability.
AF-Retriever~\cite{afretriever} prompts an LLM to translate the input query into Cypher, parses it into relational triplets and attribute constraints, grounds them on the graph, and merges the results with those of dense retrieval at a fixed ratio, all without training.
Both GraphRAFT and AF-Retriever conclude with an LLM-based reranker, fine-tuned in the former and zero-shot in the latter, that reads candidates together with their graph context.

\paragraph{Positioning of GRASP.}
GRASP is a \emph{hybrid} retriever that fuses the results of a text-based retriever and a graph-based retriever.
Combining these two has been explored before, but without optimizing how they are combined: AF-Retriever merges their results at a fixed ratio, and HybGRAG prompts an LLM to route each query to one of them.
In contrast, GRASP \emph{learns} to fuse the two retrievers.
Moreover, the existing trainable plan-based methods, MoR and GraphRAFT, train their planner solely via SFT, which requires constructing a target plan for each training query by prompting LLMs or by enumerating a few fixed templates.
Such targets are either costly to obtain in specialized domains, where LLMs struggle to synthesize accurate plans~\cite{mor}, or limited by the coverage of the templates~\cite{graphraft}.
As a result, these methods leave much of the training set unused: MoR collects 1K training queries with plans per dataset (13--17\%), while GraphRAFT retains 3.9--4.5K (56--63\%).\footnote{Computed from the released training data of GraphRAFT.}
GRASP instead warm-starts its planner with a smaller set of synthesized plans, but then explores all training queries through reinforcement learning, using rewards computed from their gold answers rather than plan annotations.

\section{Method}
\label{sec:method}
\begin{figure*}[t]
    \centering
    \includegraphics[width=\linewidth]{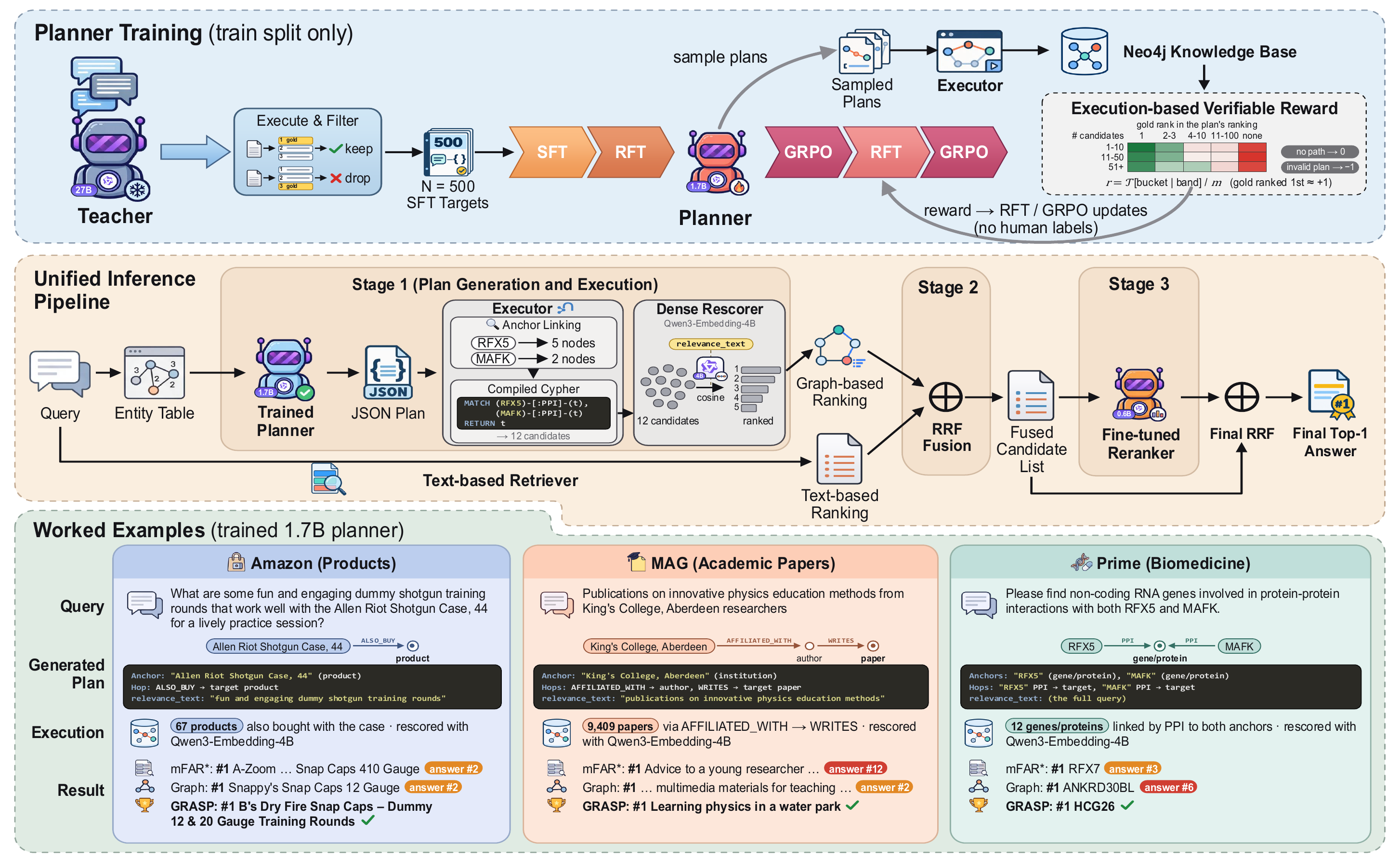}
    \caption{Overview of GRASP with concrete examples.}
    \label{fig:overview}
\end{figure*}

GRASP combines three components.
A \emph{graph branch} turns the query into a structured plan with an LLM, executes the plan against the SKB graph, and ranks the nodes it reaches.
The graph ranking is then fused with the ranking of a \emph{text branch}, i.e., a text-based retriever, and the planner is trained so that its graph ranking improves this fused list.
Finally, a trained LLM \emph{reranker} reorders the fused list.
Figure~\ref{fig:overview} gives an overview of GRASP.

\subsection{Graph branch}
\label{sec:method_stage1}

The graph branch ranks candidate nodes in four steps: it extracts the entity mentions in the query, links them to graph nodes, generates a plan, and executes the plan against the graph.
Instead of relying on multi-turn LLM interactions as in prior work \cite{avatar, afretriever}, we invoke the planner only once: it directly generates the executable plan in a single call, which avoids error propagation across turns and keeps the latency of the branch low. The rationale is that the goal of this branch is not to produce comprehensive final answers but to extract pure graph signals that complement the text signals, for which a single call suffices.

\paragraph{Entity extraction.}
A small LLM first extracts the entity mentions in the query (Figure~\ref{fig:extract-prompt}).
The prompt is one unified template, whose slots list the SKB's node types and add a few SKB-specific rules. It is followed by four demonstrations from the training set.
For each mention, the model outputs its text, written the way a graph node name would be written; its one or two most likely node types; its role; and, if the mention is nested inside a longer mention, the index of that mention.
The role is ``anchor'' if the mention names a specific entity the answer is related to, ``target'' if it describes the answer itself, and ``attribute'' if it is a constraint that is not itself a graph node (e.g., ``under \$50'').

\paragraph{Entity linking.}
Each anchor and target mention is embedded and matched against the name index of each of its predicted node types.
Because the extractor sometimes mislabels a mention, one further node type is added when its best match is near-exact (cosine${\geq} 0.99$) or beats the best match among the predicted types by at least $0.02$.
The result is an \emph{entity table} that lists every mention with its role and, for each node type, its three most similar nodes together with their cosine similarity and node degree. Attribute mentions are listed without links. An example can be found in Figure~\ref{fig:plan-prompt}.
The table gives the planner concrete node names to copy into its anchors, together with how reliable each link is (similarity) and how broad each node is (degree).

\paragraph{Plan generation.}
Given the query and its entity table, the planner emits a single JSON plan (Figure~\ref{fig:plan-prompt}).
Its system prompt is unified: it fixes the model's role, lists the SKB's node types and relations, and defines the plan format, so that only the schema is filled in per SKB.
Each plan has the form
\begin{verbatim}
{"anchors": [{"var", "text", "label",
              "match_mode"}, ...],
 "hops":    [{"from", "rel", "to_var",
              "to_label"}, ...],
 "target":  {"var": "T", "labels": [...],
             "relevance_text": ...}}
\end{verbatim}
where \texttt{var}, \texttt{from}, and \texttt{to\_var} are identifiers referring to anchors or the target; \texttt{label} and \texttt{to\_label} take values from the SKB's node types; \texttt{labels} are possible node types; and \texttt{rel} is one of the SKB's relation types.
The \texttt{match\_mode} of an anchor selects how its text is linked: \texttt{name} for head names, and \texttt{doc} for descriptive phrases.
\texttt{relevance\_text} carries the part of the query that describes the answer but cannot be expressed by hops.
% A plan may have no hops only when it has a single anchor that is itself the target. The answer is then retrieved by its description alone, which suits queries that only describe the answer.

The planner is a small LLM trained in two steps (Section~\ref{sec:method_stage2}): supervised fine-tuning on plans written by a larger teacher LLM, followed by reinforcement learning.
The teacher writes its plans from longer, dataset-specific prompts inspired by the metapath templates that STaRK provides.
%: the MAG prompt lists seven metapath templates as planning hints, the Amazon prompt adds seven planning rules and five hand-curated (query, plan) examples, and the Prime prompt appends nineteen worked examples.
The full teacher prompts are provided in our codebase.
% At inference, the planner decodes greedily. If its output cannot be parsed or fails validation, up to two further samples are drawn, and a query whose three attempts all fail receives no plan.

\paragraph{Plan execution.}
The plan is executed in four steps.
\emph{(1) Validation.}
The plan is checked against the SKB schema: every node type and relation must exist, the two node types of every hop must be connected by its relation in the SKB, and every anchor must reach the target. A plan that fails these checks returns no candidates.
\emph{(2) Anchor linking.}
Each anchor's \texttt{text} is embedded and looked up in a vector index over the nodes of its declared type. With $\texttt{match\_mode} {=} \texttt{name}$, the lookup uses an index over node names, intended for short entity strings (e.g., ``Sony''). With $\texttt{match\_mode} {=} \texttt{doc}$, it uses an index over full node documents, intended for descriptive phrases (e.g., ``headphones with long battery life''). The top-$k$ matches whose cosine similarity is at least $\tau$ times that of the best match become the anchor's candidate bindings, each carrying its cosine score (Appendix~\ref{app:impl}).
\emph{(3) Traversal.}
The plan is compiled into a single Cypher \citep{cypher} query and run on a Neo4j \citep{neo4j} backend hosting the SKB: anchors expand to \texttt{UNWIND} over their bindings, hops expand to undirected \texttt{MATCH} patterns, and every target node reached is returned with an \emph{anchor score} equal to the sum of the binding scores of the anchors on its path. When several paths reach the same target, we keep the maximum. A plan without hops skips the traversal: its candidates are the nodes of the anchor's type whose document embeddings are closest to the relevance text defined below.
\emph{(4) Dense rescoring.}
Each candidate's score is the anchor score plus the cosine similarity between the embedding of the relevance text and the candidate's document embedding, where the relevance text is \texttt{relevance\_text}, or the query itself when \texttt{relevance\_text} is empty. This handles textual constraints the hops cannot express, for instance ``at least 30 hours of battery life'' on a product already constrained to the brand ``Sony'' by a \texttt{HAS\_BRAND} hop.
The 100 highest-scoring candidates form the graph ranking. When no node satisfies the plan, the graph ranking is empty.

\subsection{Adaptive fusion with the text branch}
\label{sec:method_stage2}
Alongside the graph ranking, we run a text-based retriever over the same corpus, which produces a complementary text ranking of 100 candidates.
The two rankings are fused with reciprocal rank fusion, and the planner is trained so that its graph ranking improves the fused ranking.
The fusion weights are fixed, but the fusion is adaptive through the planner: its reward measures how much a graph ranking improves the fused ranking over the text ranking, rather than how good the graph ranking is on its own, so the graph branch learns to contribute where the text branch falls short.

\paragraph{Reciprocal rank fusion.}
Reciprocal rank fusion~\citep[RRF]{rrf} is a widely used method for fusing multiple ranked lists. Let $\mathrm{rank}_{g}(c)$ and $\mathrm{rank}_{t}(c)$ denote the ranks of candidate $c$ in the lists produced by the graph and text branches, respectively. The fused score is
\begin{equation*}
\mathrm{score}(c) = \frac{w}{k + \mathrm{rank}_{g}(c)} + \frac{1-w}{k + \mathrm{rank}_{t}(c)},
\end{equation*}
and the 100 highest-scoring candidates, in descending order of this score, form the fused ranking.
Here, $w {\in} [0,1]$ balances the contributions of the two branches, and $k {\in} \mathbb{N}$ controls how strongly top-ranked candidates are favored: a smaller $k$ amplifies differences among the top positions.
If a candidate is absent from one list, the corresponding term is set to zero. 
%In particular, when the graph branch returns no candidates, the fused ranking is exactly the text ranking.
For each dataset, we grid-search $k {\in} \{1, 2, 3, 5, 10, 20, 30, 60, 100, 200, 300\}$ and $w {\in} \{0.05, 0.10, \ldots, 0.95\}$ on the validation set with the planner after supervised fine-tuning, select the configuration with the highest Hit@1, and keep it fixed for reinforcement learning and inference.

\paragraph{Learning to fuse.}
The planner is trained to produce graph rankings that improve the fused ranking, rather than graph rankings that are good in isolation.
It is first fine-tuned on $N{=}500$ plans written by the teacher for training queries, keeping a plan only if it is valid and its executed graph ranking places a correct answer first.

Reinforcement learning (RL) then optimizes a reward that measures the gain of a plan for fusion.
Since the text ranking does not depend on the plan, a plan is worth what its graph ranking adds to the fused ranking over the text ranking alone.
We estimate this gain once, on the validation set with the fine-tuned planner and the fixed fusion configuration.
Let $u(L) = 0.7\,\mathrm{RR@10}(L) + 0.3\,\mathrm{nDCG@10}(L)$ be the utility of a ranking $L$, where $\mathrm{RR@10}$ is the reciprocal rank of the first correct answer within the top 10 (0 if there is none).
For a query, let $\Delta = u(L_{\text{fused}}) - u(L_{\text{text}})$ be the gain of fusion over the text ranking.
We bucket the number of graph candidates $n$ by $b(n) \in \{1\text{--}10,\ 11\text{--}50,\ {>}50\}$ and the rank $d$ of the first correct answer in the graph ranking by $g(d) \in \{1,\ 2\text{--}3,\ 4\text{--}10,\ 11\text{--}100,\ \varnothing\}$, where $\varnothing$ means that no correct answer is retrieved.
For each pair of buckets, $T(b, g)$ is the average $\Delta$ over the validation queries that fall into it.
The reward of a plan is
\begin{equation*}
r = \begin{cases}
T\bigl(b(n), g(d)\bigr)/m & \text{if } n \geq 1,\\
0 & \text{if } n = 0,\\
-1 & \text{if the plan is invalid},
\end{cases}
\end{equation*}
where a plan is invalid if it cannot be parsed or fails validation, and $m = \frac{1}{3}\sum_{b} T(b, 1)$ normalizes the reward so that placing a correct answer first is worth about $1$ on every dataset.
% Let $u(L) = 0.7\,\mathrm{RR@10}(L) + 0.3\,\mathrm{nDCG@10}(L)$ be the utility of a ranking $L$.
% We group validation queries by the number of graph candidates $n$ (1--10, 11--50, more than 50) and by the rank $d$ of the first correct answer in the graph ranking (1, 2--3, 4--10, 11--100, or absent), and let $T(b, d)$ be the average of $u(\text{fused}) - u(\text{text})$ over the queries in group $(b, d)$.
% The reward of a plan is
% \begin{equation*}
% r = \begin{cases}
% T\bigl(b(n), d\bigr)/m & \text{if the plan returns } n \geq 1 \text{ candidates},\\
% 0 & \text{if it returns no candidates},\\
% -1 & \text{if it is invalid or cannot be parsed},
% \end{cases}
% \end{equation*}
% where $m$ is the average of $T(b, 1)$ over the three candidate-count groups, so that placing a correct answer first is worth about $1$ on every dataset.
A plan without candidates receives $0$ because the fused ranking is then the text ranking, i.e., the plan adds nothing.
The reward only needs the executed graph ranking, so no fusion is computed during training.

RL alternates rejection-sampling fine-tuning (RFT) and GRPO \citep{grpo} for two rounds.
In each round, the current planner samples eight plans for every training query, which are executed and scored.
RFT fine-tunes the base model on the supervised data together with, for every query, its highest-reward sample if that reward is positive.
GRPO then continues from the RFT model on the training queries whose sampled rewards are not all equal.
After each round, the checkpoint with the highest fused Hit@1 on the validation set is kept.

\subsection{Reranking}
\label{sec:method_stage3}
A fine-tuned pointwise LLM reranker finally reorders the fused ranking.
Each candidate is serialized into a single text document containing (i) the node's type and its dataset-specific text fields, and (ii) one line per relation type listing up to 10 of its neighbors, chosen and ordered by the embedding similarity between each neighbor and the query.
The reranker reads each document together with the query and scores the candidate independently (prompt in Figure~\ref{fig:rerank-prompt}).

To build the fine-tuning data, we run the trained planner on the training queries and take the top 100 candidates of each fused ranking.
The correct answers among them are the positives, and the 30 highest-ranked incorrect candidates are the hard negatives. Queries without a positive are skipped.
At each training step, we build a group of $G{=}24$ candidates per query by sampling one positive and $G{-}1$ hard negatives.
We fine-tune the reranker with a cross-entropy loss over the $G$ scores that identifies the positive, which pulls the positive's score up while pushing the $G{-}1$ sampled negatives down.
Training details are in Appendix~\ref{app:impl}.

At inference, the reranker scores the top $K{=}50$ fused candidates, which are then reordered by RRF of their reranker rank and fused rank, with $(w, k)$ selected on the validation set.

\section{Experiments}
\label{sec:exp}

\subsection{Evaluation Setup}
\label{sec:exp_setup}
We evaluate on the three STaRK~\cite{stark} datasets: Amazon (product search), MAG (academic paper search), and Prime (precision-medicine inquiries). Dataset statistics are summarized in Table~\ref{tab:datasets}.
The SKBs of these datasets vary dramatically in graph density and in the relative importance of textual versus relational information. Amazon's is the most textual, with \texttt{product} nodes carrying rich content from product descriptions, customer reviews, and Q\&A data, while its relational structure is limited to a few types such as \texttt{also\_buy} and \texttt{has\_brand}. Prime's is the most relational and graph-heavy, with ten entity types and eighteen relation types spanning diseases, drugs, genes, and pathways, but with comparatively sparse text. MAG's sits in between: its \texttt{paper} nodes carry titles and abstracts, while the graph is dominated by high-cardinality \texttt{cites} and \texttt{writes} relations. 
%These differences make the three datasets a natural testbed for studying when graph structure helps over text alone. 
For evaluation, we report in percentage Hit@1 (H@1), Hit@5 (H@5), Recall@20 (R@20), and MRR. We consider text-based and graph-based baselines mentioned in Section~\ref{sec:related}, as detailed in Appendix~\ref{app:exp_baselines}.

\subsection{Implementation details}
\label{sec:exp_impl}

The entity extractor and the planner are both Qwen3-1.7B with thinking disabled.
The teacher model uses Qwen3.6-27B.
In the graph branch, anchors, relevance texts, and nodes are embedded with Qwen3-Embedding-4B.
The text branch used is mFAR \citep{mfar}, and the reranker is Qwen3-Reranker-0.6B.
Both the planner and the reranker are fine-tuned with LoRA.
Full details are given in Appendix~\ref{app:impl}.

\subsection{Main results}
\label{sec:exp_main}

\begin{table*}[t]
\centering
    \centering
    \resizebox{\ifdim\width>\textwidth \textwidth \else \width \fi}{!}{
        % 图例建议: \textbf{Bold}: best; \underline{underlined}: second best. Improvement: absolute gain over the second best.
\begin{tabular}{ll|cccc|cccc|cccc|cccc}
\toprule
\multirow{2}{*}{\textbf{Category}} & \multirow{2}{*}{\textbf{Method}} & \multicolumn{4}{c|}{\textbf{Amazon}} & \multicolumn{4}{c|}{\textbf{MAG}} & \multicolumn{4}{c|}{\textbf{Prime}} & \multicolumn{4}{c}{\textbf{Average}} \\
\cmidrule(lr){3-6} \cmidrule(lr){7-10} \cmidrule(lr){11-14} \cmidrule(lr){15-18}
& & \textbf{H@1} & \textbf{H@5} & \textbf{R@20} & \textbf{MRR} & \textbf{H@1} & \textbf{H@5} & \textbf{R@20} & \textbf{MRR} & \textbf{H@1} & \textbf{H@5} & \textbf{R@20} & \textbf{MRR} & \textbf{H@1} & \textbf{H@5} & \textbf{R@20} & \textbf{MRR} \\
\midrule
\multirow{8}{*}{\shortstack[l]{Text-\\based}}
& BM25 & 44.9 & 67.4 & 53.8 & 55.3 & 25.9 & 45.3 & 45.7 & 34.9 & 12.8 & 27.9 & 31.3 & 19.8 & 27.9 & 46.9 & 43.6 & 36.7 \\
& DPR & 15.3 & 47.9 & 44.5 & 30.2 & 10.5 & 35.2 & 42.1 & 21.3 & 4.5 & 21.9 & 30.1 & 12.4 & 10.1 & 35.0 & 38.9 & 21.3 \\
& ada-002 & 39.2 & 62.7 & 53.3 & 50.4 & 29.1 & 49.6 & 48.4 & 38.6 & 12.6 & 31.5 & 36.0 & 21.4 & 27.0 & 47.9 & 45.9 & 36.8 \\
& multi-ada-002 & 40.1 & 65.0 & 55.1 & 51.6 & 25.9 & 50.4 & 50.8 & 36.9 & 15.1 & 33.6 & 38.1 & 23.5 & 27.0 & 49.7 & 48.0 & 37.3 \\
& Qwen3 & 44.6 & 67.2 & 54.8 & 54.9 & 32.6 & 55.7 & 55.4 & 43.4 & 12.4 & 28.6 & 34.4 & 20.2 & 29.9 & 50.5 & 48.2 & 39.5 \\
& KAR & 54.2 & 68.7 & 57.2 & 61.3 & 50.5 & 69.6 & 60.3 & 58.7 & 30.4 & 49.3 & 50.8 & 39.2 & 45.0 & 62.5 & 56.1 & 53.1 \\
& mFAR & 41.2 & 70.0 & 58.5 & 54.2 & 49.0 & 69.6 & 71.7 & 58.2 & 40.9 & 62.8 & 68.3 & 51.2 & 43.7 & 67.5 & 66.2 & 54.5 \\
& mFAR$^*$ & 58.7 & \underline{82.2} & \underline{66.7} & \underline{69.3} & 53.0 & 75.2 & 74.6 & 63.0 & 39.0 & 62.6 & 67.7 & 49.7 & 50.2 & 73.3 & \underline{69.7} & 60.7 \\
\midrule
\multirow{7}{*}{\shortstack[l]{Graph-\\based}}
& QAGNN & 26.6 & 50.0 & 52.1 & 37.8 & 12.9 & 39.0 & 47.0 & 29.1 & 8.9 & 21.4 & 29.6 & 14.7 & 16.1 & 36.8 & 42.9 & 27.2 \\
& ToG & -- & -- & -- & -- & 13.2 & 16.2 & 11.3 & 14.2 & 6.1 & 15.7 & 13.1 & 10.2 & -- & -- & -- & -- \\
& AvaTaR & 49.9 & 69.2 & 60.6 & 58.7 & 44.4 & 59.7 & 50.6 & 51.2 & 18.4 & 36.7 & 39.3 & 26.7 & 37.6 & 55.2 & 50.2 & 45.5 \\
& HybGRAG & -- & -- & -- & -- & 65.4 & 75.3 & 65.7 & 69.8 & 28.6 & 41.4 & 43.6 & 34.5 & -- & -- & -- & -- \\
& MoR & 52.2 & 74.7 & 59.9 & 62.2 & 58.2 & 78.3 & 75.0 & 67.1 & 36.4 & 60.0 & 63.5 & 46.9 & 48.9 & 71.0 & 66.1 & 58.7 \\
& GraphRAFT & -- & -- & -- & -- & 69.6 & 84.3 & \underline{89.1} & 76.2 & \underline{63.7} & \underline{75.4} & \underline{76.4} & \underline{69.0} & -- & -- & -- & -- \\
& AF-Retriever & \underline{61.2} & 75.2 & 35.5 & 67.3 & \underline{78.6} & \underline{91.4} & 61.4 & \underline{84.0} & 46.2 & 63.7 & 51.2 & 54.0 & \underline{62.0} & \underline{76.8} & 49.4 & \underline{68.4} \\
\midrule
Hybrid & \textbf{GRASP (ours)} & \textbf{75.5} & \textbf{89.3} & \textbf{74.1} & \textbf{81.8} & \textbf{89.1} & \textbf{95.8} & \textbf{94.3} & \textbf{92.1} & \textbf{77.3} & \textbf{92.1} & \textbf{92.0} & \textbf{83.7} & \textbf{80.6} & \textbf{92.4} & \textbf{86.8} & \textbf{85.9} \\
& \textit{Improvement} & \textit{+14.3} & \textit{+7.1} & \textit{+7.4} & \textit{+12.5} & \textit{+10.5} & \textit{+4.4} & \textit{+5.2} & \textit{+8.1} & \textit{+13.6} & \textit{+16.7} & \textit{+15.6} & \textit{+14.7} & \textit{+18.6} & \textit{+15.6} & \textit{+17.1} & \textit{+17.5} \\
\bottomrule
\end{tabular}
    }
    \caption{Main results on Amazon, MAG, and Prime.
    Our method, GRASP, achieves the best performance on all metrics across all three datasets.
    \textbf{Bold} indicates the best result and \underline{underline} the second best, per metric per dataset.
    The \textit{Improvement} row reports GRASP's absolute improvement over the second-best result. mFAR$^*$ denotes our re-run of mFAR. ``--'' denotes numbers not reported.}
    % mFAR uses 1-hop relational information only as additional text fields and is thus grouped with text-based methods.
    \label{tab:main_results}
\end{table*}

Table~\ref{tab:main_results} compares GRASP with text-based and graph-based baselines.
GRASP achieves the best performance on every metric across all three datasets.
On average, it reaches 80.6 H@1, outperforming the strongest baseline, AF-Retriever, by 18.6 points.
The gains are consistent across datasets, with ${>}10$ H@1 improvements over the second-best result on Amazon, MAG, and Prime.
Table~\ref{tab:human_eval_results} reports results on the human-generated subset.
GRASP again leads by a substantial margin of 11.4 average H@1, reaching 81.3, close to 80.6 on the full test set, indicating that its gains carry over to natural human queries.
Table~\ref{tab:stage_latency} breaks down the per-query latency of GRASP.
The full pipeline takes 2.47\,s per query on average.
The reranker accounts for 1.12\,s (45\%), and most of the remaining time is spent in the graph branch, mainly on plan generation (0.53\,s) and entity linking (0.46\,s).
Plan execution takes only 0.12\,s, fusion is negligible, and the text branch runs in parallel with the graph branch, so it adds no latency.
Latency can be further traded for accuracy through the reranker (Table~\ref{tab:ablation_branch}).
Reducing $K$ to 20 cuts latency to 1.72\,s with an average H@1 of 80.1, and removing the reranker cuts it to 1.35\,s while still reaching 67.7 average H@1, higher than every baseline in Table~\ref{tab:main_results}. On the other hand, the strongest baseline, AF-Retriever, has a latency of 36.4\,s \cite{afretriever}, far slower than GRASP.

Hyperparameter sensitivity analysis is conducted in Appendix~\ref{app:exp_sensitivity}, showing that GRASP is not sensitive to the hyperparameters used in the RRF fusions.

\subsection{Ablation study}
\label{sec:exp_ablation}

\begin{table*}[t]
\centering
    \centering
    \resizebox{\ifdim\width>\textwidth \textwidth \else \width \fi}{!}{
        \begin{tabular}{l|ccccc|ccccc|ccccc|ccccc}
\toprule
\multirow{2}{*}{\textbf{Variant}} & \multicolumn{5}{c|}{\textbf{Amazon}} & \multicolumn{5}{c|}{\textbf{MAG}} & \multicolumn{5}{c|}{\textbf{Prime}} & \multicolumn{5}{c}{\textbf{Average}} \\
\cmidrule(lr){2-6} \cmidrule(lr){7-11} \cmidrule(lr){12-16} \cmidrule(lr){17-21}
& \textbf{H@1} & \textbf{H@5} & \textbf{R@20} & \textbf{R@50} & \textbf{MRR} & \textbf{H@1} & \textbf{H@5} & \textbf{R@20} & \textbf{R@50} & \textbf{MRR} & \textbf{H@1} & \textbf{H@5} & \textbf{R@20} & \textbf{R@50} & \textbf{MRR} & \textbf{H@1} & \textbf{H@5} & \textbf{R@20} & \textbf{R@50} & \textbf{MRR} \\
\midrule
\multicolumn{21}{l}{\textit{w/ reranker ($K{=}50$, latency: 2.47\,s/query)}} \\
\midrule
GRASP (full) & \textbf{75.5} & \textbf{89.3} & \textbf{74.1} & \textbf{79.0} & \textbf{81.8} & \textbf{89.1} & \textbf{95.8} & \textbf{94.3} & \textbf{96.4} & \textbf{92.1} & \textbf{77.3} & \textbf{92.1} & \textbf{92.0} & \textbf{93.0} & \textbf{83.7} & \textbf{80.6} & \textbf{92.4} & \textbf{86.8} & \textbf{89.5} & \textbf{85.9} \\
\quad Graph only & 60.9\rlap{$^*$} & 72.5 & 55.2 & 57.9 & 66.1 & 88.8 & 94.1 & 92.9 & 94.3 & 91.2 & 66.5\rlap{$^*$} & 76.4 & 75.6 & 76.4 & 70.8 & 72.1\rlap{$^*$} & 81.0 & 74.6 & 76.2 & 76.0 \\
\quad Text only & \textbf{75.5} & 89.0 & 72.0 & 76.7 & 81.5 & 72.3\rlap{$^*$} & 84.5 & 78.9 & 80.4 & 77.8 & 65.4\rlap{$^*$} & 77.9 & 75.0 & 75.6 & 70.9 & 71.1\rlap{$^*$} & 83.8 & 75.3 & 77.6 & 76.7 \\
\midrule
\multicolumn{21}{l}{\textit{w/ reranker ($K{=}20$, latency: 1.72\,s/query)}} \\
\midrule
GRASP & \textbf{75.2} & \textbf{88.7} & \textbf{69.7} & \textbf{79.0} & \textbf{81.3} & \textbf{88.8} & \textbf{95.1} & \textbf{92.7} & \textbf{96.4} & \textbf{91.6} & \textbf{76.2} & \textbf{90.0} & \textbf{87.8} & \textbf{93.0} & \textbf{82.4} & \textbf{80.1} & \textbf{91.3} & \textbf{83.4} & \textbf{89.5} & \textbf{85.1} \\
\quad Graph only & 58.6\rlap{$^*$} & 69.1 & 49.9 & 57.9 & 63.5 & 88.5 & 94.0 & 91.4 & 94.3 & 90.9 & 65.8\rlap{$^*$} & 75.3 & 73.7 & 76.4 & 70.0 & 71.0\rlap{$^*$} & 79.5 & 71.7 & 76.2 & 74.8 \\
\quad Text only & 74.9 & 88.5 & 66.7 & 76.7 & 80.8 & 71.5\rlap{$^*$} & 81.9 & 74.6 & 80.4 & 76.2 & 62.0\rlap{$^*$} & 72.7 & 67.7 & 75.6 & 66.9 & 69.5\rlap{$^*$} & 81.0 & 69.7 & 77.6 & 74.6 \\
\midrule
\multicolumn{21}{l}{\textit{w/o reranker (latency: 1.35\,s/query)}} \\
\midrule
GRASP & \textbf{60.5} & \textbf{84.5} & \textbf{69.7} & \textbf{79.0} & \textbf{71.3} & \textbf{80.7} & \textbf{91.9} & \textbf{92.7} & \textbf{96.4} & \textbf{85.8} & \textbf{62.0} & \textbf{83.6} & \textbf{87.8} & \textbf{93.0} & \textbf{71.7} & \textbf{67.7} & \textbf{86.7} & \textbf{83.4} & \textbf{89.5} & \textbf{76.3} \\
\quad Graph only & 41.2\rlap{$^*$} & 60.7 & 49.9 & 57.9 & 50.1 & 80.2 & 91.7 & 91.4 & 94.3 & 85.4 & 55.4\rlap{$^*$} & 72.0 & 73.7 & 76.4 & 62.6 & 58.9\rlap{$^*$} & 74.8 & 71.7 & 76.2 & 66.0 \\
\quad Text only & 58.7\rlap{$^*$} & 82.2 & 66.7 & 76.7 & 69.3 & 53.0\rlap{$^*$} & 75.2 & 74.6 & 80.4 & 63.0 & 39.0\rlap{$^*$} & 62.6 & 67.7 & 75.6 & 49.7 & 50.2\rlap{$^*$} & 73.3 & 69.7 & 77.6 & 60.7 \\
\bottomrule
\end{tabular}

    }
    \caption{Ablation on GRASP components.
    Graph only and Text only use a single branch.
    Latency is the average per-query time of GRASP in each setting.
    \textbf{Bold} indicates the best result per metric per dataset within each setting. $^*$ marks H@1 results that differ significantly from GRASP in the same setting ($p<0.05$, McNemar's test, Stouffer's method for Average).}
    \label{tab:ablation_branch}
\end{table*}

We ablate GRASP along three axes, namely the two branches and the reranker (Section~\ref{sec:ablation_branch}), the design and training of the graph branch (Section~\ref{sec:ablation_graph}), and the choice of text retriever (Section~\ref{sec:ablation_text}).
Each ablation is evaluated both with and without the reranker, since the reranker can mask differences in retrieval quality.
 
\subsubsection{Branches and reranker}
\label{sec:ablation_branch}
 
\paragraph{Both branches are necessary.}
Table~\ref{tab:ablation_branch} shows that removing either branch significantly lowers the average H@1 in every setting, with or without the reranker.
% With the reranker, Graph only and Text only reach 72.1 and 71.1, compared with 80.6 for GRASP.
% Without the reranker, they reach 58.9 and 50.2, compared with 67.7.
The two branches are also complementary, and their strengths again follow the characteristics of each SKB (Section~\ref{sec:exp_setup}).
Without the reranker, the graph branch outperforms the text branch by 27.2 H@1 points on MAG and 16.4 points on Prime (both are relation-heavier), whereas the text branch leads by 17.5 points on Amazon (text-heavier).
Combining them yields the largest gain on Prime, where GRASP improves over the better single branch by 10.8 points with the reranker, suggesting that the two branches retrieve complementary candidates.
On MAG and Amazon, where one branch dominates, GRASP matches the stronger branch, so fusion does not hurt. 
These results show that the fusion of the graph and text branches makes GRASP robust to various SKBs. 
 
\paragraph{Reranker.}
The reranker substantially improves ranking quality.
For GRASP, it raises the average H@1 from 67.7 to 80.6 and MRR from 76.3 to 85.9, at a cost of 1.12\,s per query.
Reducing $K$ from 50 to 20 lowers latency from 2.47\,s to 1.72\,s with only a small drop of 0.5 in average H@1.
% R@20, however, falls back to its value without the reranker (83.4), because reordering only the top 20 candidates cannot change which candidates are in the top 20.

\subsubsection{Graph branch}
\label{sec:ablation_graph}

We first examine two design choices of the graph branch (Table~\ref{tab:ablation_graph_design}) and then study how it should be trained (Table~\ref{tab:ablation_graph_training}).
 
\paragraph{Entity table.}
The entity table lists the entity mentions extracted from the query, each with its most similar graph nodes, their similarity, and their degree, and is included in the prompt of the planner (Section~\ref{sec:method_stage1}).
Removing it significantly lowers the average H@1 without the reranker, and the drop comes almost entirely from Prime.
Prime has the richest schema, so grounding query mentions to the right nodes is hardest there, and the pre-linked candidates help the planner choose correct anchors.
% On Amazon and MAG, H@1 changes by at most 0.5 points.
% With the reranker, the difference disappears (80.6 for both).
 
\paragraph{Relevance text.}
The relevance text is the part of the query that describes the answer but is not captured by the hops, and candidates are rescored by their document similarity to it.
Without it, candidates are rescored against the full query instead.
This significantly lowers the average H@1 without the reranker (67.7 to 64.8), almost entirely due to MAG (80.7 to 71.9), while Amazon and Prime are essentially unchanged.
Many MAG queries combine a relational constraint, such as an author or a cited paper, with a topical description of the target paper.
Because the cites and writes relations have high cardinality, traversal often returns many candidates that must be ranked by their titles and abstracts.
The full query mixes the relational constraint, which the hops already enforce, with the topical description, whereas the relevance text isolates the latter and therefore ranks these candidates more accurately.
With the reranker, which scores candidates against the full query, the difference largely disappears.
We discuss the unchanged result on Amazon in Appendix~\ref{app:nothelpamazon}.
 
\paragraph{Training setup.}
We study how the graph branch should be trained by varying two factors.
First, we compare SFT on teacher-written plans alone with SFT followed by RL, which alternates RFT and GRPO for two rounds (Section~\ref{sec:method_stage2}).
Second, we vary the amount of SFT data, using 1/4, 1/2, and the full SFT set ($N{=}500$), and apply RL on top of each.
We also include the teacher model that generates the SFT data as a reference point for distillation.
Finally, we break down the effect of RL by plan depth to understand where the gains come from.
 
\paragraph{Effect of RL and SFT data.}
Table~\ref{tab:ablation_graph_training} and Figure~\ref{fig:sft_scaling} show that RL consistently improves the graph branch at every SFT data size.
Without the reranker, RL raises the average H@1 by 5.9, 5.7, and 4.4 points with 1/4, 1/2, and the full SFT set, respectively, and the corresponding gains with the reranker are 3.9, 2.3, and 1.5 points.
The gain is largest when SFT data is scarce, suggesting that RL compensates for limited supervision.
RL is also more data-efficient than additional SFT data.
With only half of the SFT set, SFT + RL already outperforms SFT on the full set (66.5 vs.\ 63.3 without the reranker and 80.0 vs.\ 79.1 with it).
Meanwhile, the benefit of more SFT data diminishes quickly, from +5.4 points when doubling 125 to 250 examples to +2.5 points when doubling 250 to 500, indicating that 500 examples are close to saturation.
 
\paragraph{Surpassing the teacher.}
SFT alone never reaches the teacher, even with the full SFT set (63.3 vs.\ 64.2 without the reranker and 79.1 vs.\ 79.6 with it), since imitating teacher-generated plans bounds the student by the teacher.
With RL, the 1.7B student surpasses the 27B teacher from $N{=}250$ onward, and the final model outperforms it by 3.5 and 1.0 points in average H@1 without and with the reranker, both statistically significant.
The teacher remains stronger on the recall-oriented metrics of MAG (e.g., 97.3 vs.\ 96.4 R@50 without the reranker), but its H@1 is significantly lower (78.9 vs.\ 80.7), showing that RL particularly sharpens the top-ranked prediction.
 
\paragraph{Where the gains come from.}
Figure~\ref{fig:hop_analysis} breaks down the results without the reranker by plan depth, with full numbers in Table~\ref{tab:hop_analysis}.
The gains are concentrated on Prime, where H@1 rises from 50.4 to 62.0.
RL shifts Prime plans toward deeper traversals, increasing the share of plans with two or more hops from 35.4\% to 54.0\%, and reduces failed plan generation from 6.9\% to 1.4\%.
At the same time, accuracy improves at every depth (from 46.2 to 56.7 for one hop, 57.6 to 68.9 for two hops, and 66.2 to 73.6 for three or more hops).
Holding the depth distribution fixed, the within-depth improvement accounts for about 8.7 points of the total gain, and the remaining 3.0 points come from the shift toward deeper plans.
On MAG, the depth distribution barely changes (56.2\% vs.\ 56.8\% multi-hop plans), and the modest gain of 1.3 points comes mainly from better one-hop and three-or-more-hop plans.

\subsubsection{Text branch}
\label{sec:ablation_text}
 
Table~\ref{tab:ablation_text} replaces mFAR with Qwen3 in the text branch.
The average H@1 drops significantly in both settings (80.6 to 77.0 with the reranker and 67.7 to 61.0 without it).
The drop is concentrated on Amazon (75.5 to 70.5 with the reranker) and Prime (77.3 to 72.4), while MAG changes little (89.1 to 88.2), consistent with MAG relying mainly on the graph branch (Section~\ref{sec:ablation_branch}).
Recall drops as well, with R@50 falling from 79.0 to 70.8 on Amazon and from 93.0 to 85.3 on Prime, showing that a weaker text branch supplies fewer correct candidates, which the reranker cannot recover.
Nevertheless, GRASP with Qwen3 still outperforms every baseline in Table~\ref{tab:main_results} (77.0 vs.\ 62.0 average H@1), and even without the reranker it more than doubles the H@1 of Qwen3 alone (61.0 vs.\ 29.9).
The benefit of GRASP therefore does not hinge on a particular text retriever.

\section{Conclusion}
\label{sec:conclusion}
We presented GRASP, a hybrid retriever for SKBs that combines a plan-guided graph branch with a text branch and a reranker.
Instead of training the graph retriever in isolation, GRASP trains its planner with a reward that measures how much the graph ranking improves the fused ranking over the text ranking, so that the graph branch learns to complement the text branch.
On the three STaRK benchmarks, GRASP outperforms both text-based and graph-based baselines on all metrics while answering a query in a few seconds.
Our ablations show that both branches and the reranker are necessary, and that RL lets a small planner surpass its much larger teacher, with the largest gains on relation-heavy SKBs where it learns deeper and more accurate plans.
We hope that learning to fuse, rather than fusing independently trained retrievers, will benefit retrieval over other structured and semi-structured data.

\section*{Limitations}
\label{sec:limitations}
\paragraph{Generalization beyond STaRK.} Our experiments cover three SKB benchmarks (Amazon, MAG, Prime) that share the STaRK construction protocol and a common English schema. Transfer to other SKBs (e.g., industrial product catalogs, non-English biomedical KGs, or KGs with substantially deeper relation hierarchies) remains untested; we expect plan generation to be the component most sensitive to such shifts, since it relies on the LLM recognizing the dataset schema in-context.

\paragraph{Interpretability.} While our ablations attribute performance to specific components, we do not provide a systematic account of \emph{why} a given top-$1$ prediction emerges from the interaction of different components. Mapping each retrieval decision back to a chain of (graph branch, text branch, reranker) interactions is left to future work.

%
% \section*{Acknowledgments}

\bibliography{custom}

\appendix
\renewcommand{\thetable}{A\arabic{table}}
\renewcommand{\thefigure}{A\arabic{figure}}
\setcounter{table}{0}
\setcounter{figure}{0}

\begin{figure}
    \centering
    \includegraphics[width=\columnwidth]{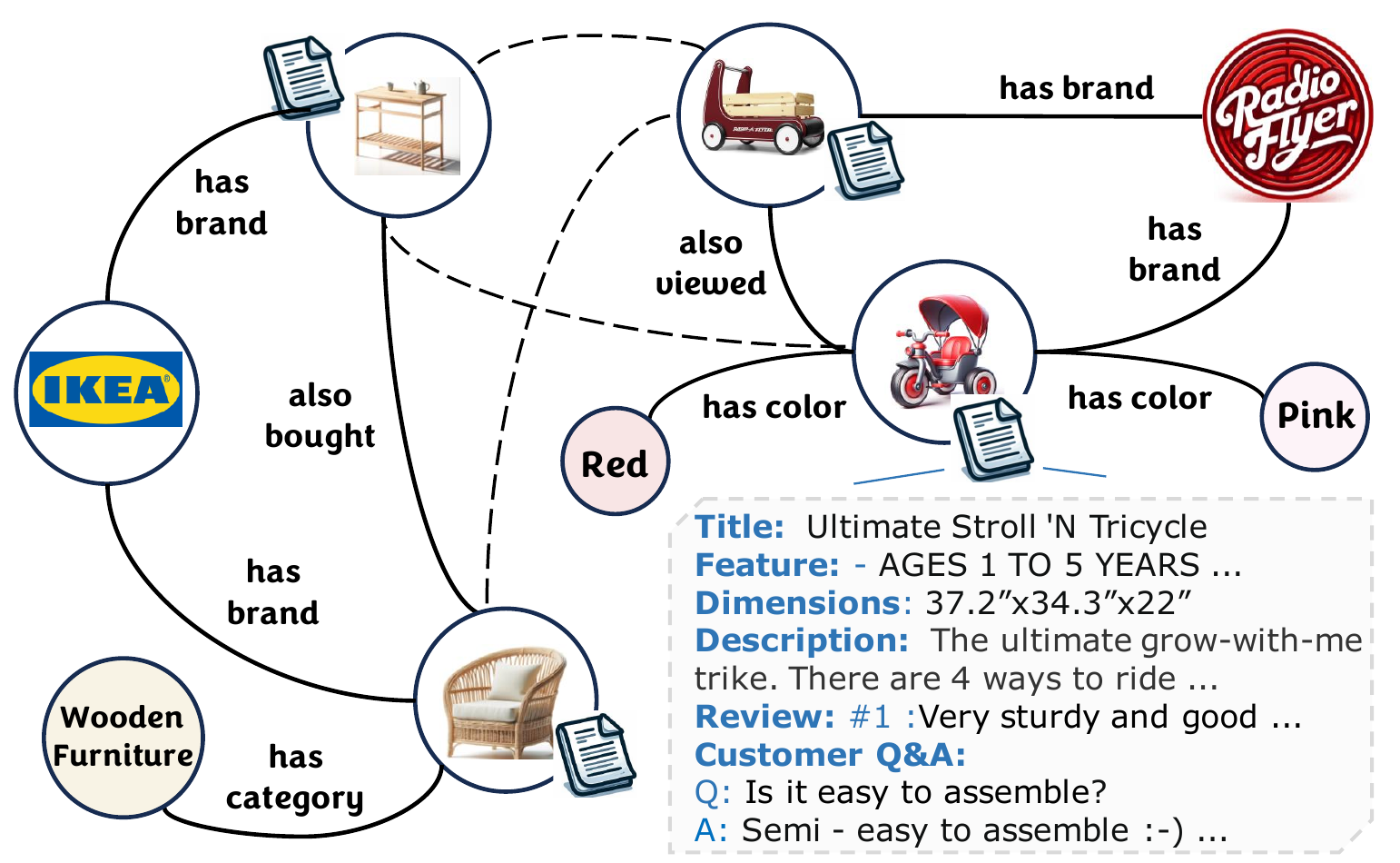}
    \caption{An example of Amazon SKB reproduced from \citet{stark}. The attributed node shown here is of type ``product'', with the ``Title'' attribute as its head name. The relations connecting this node include ``has color'', ``has brand'', and ``also viewed''.}
    \label{fig:amazon_exp}
\end{figure}

\begin{table*}[t]
    \centering
    \resizebox{\ifdim\width>\textwidth \textwidth \else \width \fi}{!}{
        \begin{tabular}{lrrrrrrrc}
\toprule
\textbf{Dataset} & \textbf{Nodes} & \textbf{Relations} & \textbf{Avg. deg.} & \textbf{Node types} & \textbf{Rel.\ types} & \textbf{Tokens} & \textbf{Queries} & \textbf{Train / Val / Test (\%)} \\
\midrule
Amazon  & 1.04M & 9.44M  & 18.2  & 4  & 5  & 592.1M & 9{,}100  & 65 / 17 / 18 \\
MAG     & 1.87M & 39.8M  & 43.5  & 4  & 4  & 212.6M & 13{,}300 & 60 / 20 / 20 \\
Prime   & 0.13M & 8.10M  & 125.2 & 10 & 18 & 31.8M  & 11{,}200 & 55 / 20 / 25 \\
\bottomrule
\end{tabular}
    }
    \caption{STaRK dataset statistics.}
    \label{tab:datasets}
\end{table*}

\begin{table*}[t]
    \centering
    \resizebox{\ifdim\width>\textwidth \textwidth \else \width \fi}{!}{
        \begin{tabular}{ll|cccc|cccc|cccc|cccc}
\toprule
\multirow{2}{*}{\textbf{Category}} & \multirow{2}{*}{\textbf{Method}} & \multicolumn{4}{c|}{\textbf{Amazon}} & \multicolumn{4}{c|}{\textbf{MAG}} & \multicolumn{4}{c|}{\textbf{Prime}} & \multicolumn{4}{c}{\textbf{Average}} \\
\cmidrule(lr){3-6} \cmidrule(lr){7-10} \cmidrule(lr){11-14} \cmidrule(lr){15-18}
& & \textbf{H@1} & \textbf{H@5} & \textbf{R@20} & \textbf{MRR} & \textbf{H@1} & \textbf{H@5} & \textbf{R@20} & \textbf{MRR} & \textbf{H@1} & \textbf{H@5} & \textbf{R@20} & \textbf{MRR} & \textbf{H@1} & \textbf{H@5} & \textbf{R@20} & \textbf{MRR} \\
\midrule
\multirow{7}{*}{\shortstack[l]{Text-\\based}}
& BM25 & 27.2 & 51.9 & 29.2 & 18.8 & 32.1 & 41.7 & 32.5 & 37.4 & 22.5 & 41.8 & 42.3 & 30.4 & 27.3 & 45.1 & 34.7 & 28.9 \\
& DPR & 16.1 & 39.5 & 15.2 & 27.2 & 4.7 & 9.5 & 25.0 & 7.9 & 2.0 & 9.2 & 10.7 & 7.1 & 7.6 & 19.4 & 17.0 & 14.1 \\
& ada-002 & 39.5 & 64.2 & 35.5 & 52.7 & 28.6 & 41.7 & 36.0 & 35.8 & 17.4 & 34.7 & 41.1 & 26.4 & 28.5 & 46.9 & 37.5 & 38.3 \\
& multi-ada-002 & 46.9 & 72.8 & 40.2 & 58.7 & 23.8 & 41.7 & 39.9 & 31.4 & 24.5 & 39.8 & 47.2 & 33.0 & 31.7 & 51.4 & 42.4 & 41.0 \\
& Qwen3 & 43.9 & 70.7 & 56.0 & 55.2 & 35.3 & 56.4 & 56.6 & 45.5 & 14.6 & 27.1 & 34.8 & 21.0 & 31.3 & 51.4 & 49.1 & 40.6 \\
& KAR & 61.7 & 72.8 & 40.6 & 66.3 & 51.2 & 58.3 & 46.6 & 54.5 & 45.0 & 60.6 & 59.9 & 51.9 & 52.6 & 63.9 & 49.0 & 57.6 \\
& mFAR$^*$ & \underline{75.6} & \textbf{90.9} & \underline{72.3} & \underline{82.2} & \underline{69.9} & \underline{83.1} & \underline{74.8} & \underline{75.3} & \underline{64.3} & \underline{75.0} & \underline{73.7} & \underline{68.8} & \underline{69.9} & \underline{83.0} & \underline{73.6} & \underline{75.4} \\
\midrule
\multirow{3}{*}{\shortstack[l]{Graph-\\based}}
& QAGNN & 22.2 & 49.4 & 21.5 & 31.3 & 20.2 & 26.2 & 28.8 & 25.5 & 6.1 & 13.3 & 17.6 & 9.4 & 16.2 & 29.6 & 22.6 & 22.1 \\
& AvaTaR & 58.0 & \underline{76.5} & -- & 65.9 & 33.3 & 42.9 & -- & 38.6 & 33.0 & 51.4 & -- & 41.0 & 41.4 & 56.9 & -- & 48.5 \\
& AF-Retriever & 58.0 & 69.1 & -- & 63.5 & 52.4 & 60.7 & -- & 55.9 & 57.1 & 69.4 & -- & 62.7 & 55.8 & 66.4 & -- & 60.7 \\
\midrule
Hybrid & \textbf{GRASP (ours)} & \textbf{76.2} & \textbf{90.9} & \textbf{73.9} & \textbf{82.8} & \textbf{90.2} & \textbf{97.0} & \textbf{94.3} & \textbf{93.0} & \textbf{77.5} & \textbf{92.1} & \textbf{92.2} & \textbf{83.6} & \textbf{81.3} & \textbf{93.3} & \textbf{86.8} & \textbf{86.5} \\
& \textit{Improvement} & \textit{+0.6} & \textit{+0.0} & \textit{+1.6} & \textit{+0.6} & \textit{+20.3} & \textit{+13.9} & \textit{+19.5} & \textit{+17.7} & \textit{+13.2} & \textit{+17.1} & \textit{+18.5} & \textit{+14.8} & \textit{+11.4} & \textit{+10.3} & \textit{+13.2} & \textit{+11.1} \\
\bottomrule
\end{tabular}
    }
    \caption{Results on the human-generated subset across three datasets, including only baselines with available results.
    GRASP achieves the best or tied-best results on all metrics, with performance consistent with that on the full test set (Table~\ref{tab:main_results}).
    Notations follow Table~\ref{tab:main_results}.}
    \label{tab:human_eval_results}
\end{table*}

\begin{table}[t]
    \centering
    \resizebox{\ifdim\width>\columnwidth \columnwidth \else \width \fi}{!}{
        \begin{tabular}{l|cccc|c|c|c}
\toprule
\multirow{2}{*}[-2ex]{\begin{tabular}[c]{@{}l@{}}\textbf{Latency}\\\textbf{(s/query)}\end{tabular}} & \multicolumn{4}{c|}{\textbf{Graph branch} (text branch in parallel)} & \multirow{2}{*}[-2ex]{\textbf{Fusion}} & \multirow{2}{*}[-2ex]{\textbf{Reranker}} & \multirow{2}{*}[-2ex]{\textbf{Total}} \\
\cmidrule(lr){2-5}
& \begin{tabular}[c]{@{}c@{}}\textbf{Entity}\\\textbf{extraction}\end{tabular} & \begin{tabular}[c]{@{}c@{}}\textbf{Entity}\\\textbf{linking}\end{tabular} & \begin{tabular}[c]{@{}c@{}}\textbf{Plan}\\\textbf{generation}\end{tabular} & \begin{tabular}[c]{@{}c@{}}\textbf{Plan}\\\textbf{execution}\end{tabular} & & & \\
\midrule
GRASP (full) & 0.25 & 0.46 & 0.53 & 0.12 & $<$0.01 & 1.12 & 2.47 \\
\bottomrule
\end{tabular}
    }
    \caption{Per-query latency breakdown of GRASP on average.
    The text branch runs in parallel with the graph branch, so its latency is fully overlapped.}
    \label{tab:stage_latency}
\end{table}

\begin{table*}[t]
\centering
    \centering
    \resizebox{\ifdim\width>\textwidth \textwidth \else \width \fi}{!}{
        \begin{tabular}{l|ccccc|ccccc|ccccc|ccccc}
\toprule
\multirow{2}{*}{\textbf{Variant}} & \multicolumn{5}{c|}{\textbf{Amazon}} & \multicolumn{5}{c|}{\textbf{MAG}} & \multicolumn{5}{c|}{\textbf{Prime}} & \multicolumn{5}{c}{\textbf{Average}} \\
\cmidrule(lr){2-6} \cmidrule(lr){7-11} \cmidrule(lr){12-16} \cmidrule(lr){17-21}
& \textbf{H@1} & \textbf{H@5} & \textbf{R@20} & \textbf{R@50} & \textbf{MRR} & \textbf{H@1} & \textbf{H@5} & \textbf{R@20} & \textbf{R@50} & \textbf{MRR} & \textbf{H@1} & \textbf{H@5} & \textbf{R@20} & \textbf{R@50} & \textbf{MRR} & \textbf{H@1} & \textbf{H@5} & \textbf{R@20} & \textbf{R@50} & \textbf{MRR} \\
\midrule
\multicolumn{21}{l}{\textit{w/ reranker ($K{=}50$)}} \\
\midrule
GRASP (full) & 75.5 & 89.3 & 74.1 & 79.0 & 81.8 & 89.1 & 95.8 & 94.3 & 96.4 & 92.1 & \textbf{77.3} & \textbf{92.1} & \textbf{92.0} & \textbf{93.0} & \textbf{83.7} & \textbf{80.6} & 92.4 & \textbf{86.8} & 89.5 & \textbf{85.9} \\
\quad w/o entity table & \textbf{75.9} & 89.6 & 74.2 & 79.2 & 81.9 & \textbf{89.2} & \textbf{96.2} & \textbf{94.5} & 96.3 & \textbf{92.2} & 76.8 & \textbf{92.1} & 91.6 & 92.6 & 83.5 & \textbf{80.6} & 92.6 & \textbf{86.8} & 89.4 & \textbf{85.9} \\
\quad w/o relevance text & 75.6 & \textbf{90.7} & \textbf{74.4} & \textbf{79.6} & \textbf{82.1} & 88.8 & 95.8 & 93.8 & \textbf{96.5} & 91.9 & 77.0 & 91.8 & \textbf{92.0} & \textbf{93.0} & 83.6 & 80.5 & \textbf{92.8} & 86.7 & \textbf{89.7} & \textbf{85.9} \\
\midrule
\multicolumn{21}{l}{\textit{w/o reranker}} \\
\midrule
GRASP & 60.5 & \textbf{84.5} & 69.7 & 79.0 & 71.3 & \textbf{80.7} & \textbf{91.9} & \textbf{92.7} & 96.4 & \textbf{85.8} & \textbf{62.0} & \textbf{83.6} & 87.8 & \textbf{93.0} & \textbf{71.7} & \textbf{67.7} & \textbf{86.7} & 83.4 & 89.5 & \textbf{76.3} \\
\quad w/o entity table & \textbf{60.9} & \textbf{84.5} & 69.7 & 79.2 & \textbf{71.4} & 80.2 & 91.3 & 92.1 & 96.3 & 85.3 & 59.5\rlap{$^*$} & 82.8 & 87.4 & 92.6 & 69.7 & 66.9\rlap{$^*$} & 86.2 & 83.1 & 89.4 & 75.5 \\
\quad w/o relevance text & 60.5 & \textbf{84.5} & \textbf{70.1} & \textbf{79.6} & 71.2 & 71.9\rlap{$^*$} & 90.9 & 92.5 & \textbf{96.5} & 80.4 & \textbf{62.0} & \textbf{83.6} & \textbf{87.9} & \textbf{93.0} & \textbf{71.7} & 64.8\rlap{$^*$} & 86.3 & \textbf{83.5} & \textbf{89.7} & 74.4 \\
\bottomrule
\end{tabular}

    }
    \caption{Ablation on the graph branch design. \textbf{Bold} indicates the best result per metric per dataset within each setting. $^*$ marks H@1 results that differ significantly from GRASP in the same setting ($p<0.05$, McNemar's test, Stouffer's method for Average).}
    \label{tab:ablation_graph_design}
\end{table*}

\begin{figure}[t]
    \centering
    \includegraphics[width=\columnwidth]{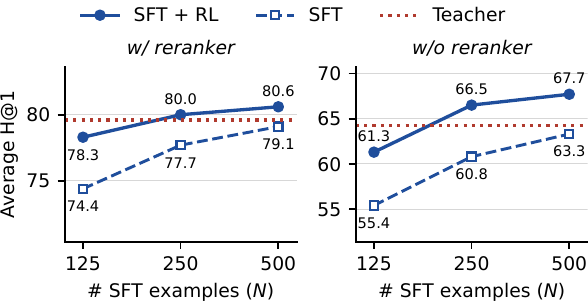}
    \caption{Average H@1 of GRASP with different amounts of SFT data, with and without the reranker.
    Solid and dashed lines denote SFT + RL and SFT only, and the dotted line denotes the teacher model.}
    \label{fig:sft_scaling}
\end{figure}

\begin{table*}[t]
\centering
    \centering
    \resizebox{\ifdim\width>\textwidth \textwidth \else \width \fi}{!}{
        \begin{tabular}{l|ccccc|ccccc|ccccc|ccccc}
\toprule
\multirow{2}{*}{\textbf{Variant}} & \multicolumn{5}{c|}{\textbf{Amazon}} & \multicolumn{5}{c|}{\textbf{MAG}} & \multicolumn{5}{c|}{\textbf{Prime}} & \multicolumn{5}{c}{\textbf{Average}} \\
\cmidrule(lr){2-6} \cmidrule(lr){7-11} \cmidrule(lr){12-16} \cmidrule(lr){17-21}
& \textbf{H@1} & \textbf{H@5} & \textbf{R@20} & \textbf{R@50} & \textbf{MRR} & \textbf{H@1} & \textbf{H@5} & \textbf{R@20} & \textbf{R@50} & \textbf{MRR} & \textbf{H@1} & \textbf{H@5} & \textbf{R@20} & \textbf{R@50} & \textbf{MRR} & \textbf{H@1} & \textbf{H@5} & \textbf{R@20} & \textbf{R@50} & \textbf{MRR} \\
\midrule
\multicolumn{21}{l}{\textit{w/ reranker ($K{=}50$)}} \\
\midrule
GRASP (full, SFT + RL) & 75.5 & 89.3 & \textbf{74.1} & \textbf{79.0} & 81.8 & \textbf{89.1} & 95.8 & 94.3 & 96.4 & 92.1 & \textbf{77.3} & \textbf{92.1} & \textbf{92.0} & \textbf{93.0} & \textbf{83.7} & \textbf{80.6} & \textbf{92.4} & \textbf{86.8} & \textbf{89.5} & \textbf{85.9} \\
\quad SFT (1/4 data) & 74.7 & 89.0 & 72.7 & 77.1 & 81.2 & 80.0\rlap{$^*$} & 91.5 & 88.0 & 90.2 & 85.1 & 68.5\rlap{$^*$} & 82.0 & 79.9 & 80.6 & 74.5 & 74.4\rlap{$^*$} & 87.5 & 80.2 & 82.6 & 80.3 \\
\quad SFT (1/2 data) & 75.3 & 89.3 & 73.7 & 78.2 & 81.6 & 86.7\rlap{$^*$} & 94.0 & 92.4 & 94.1 & 89.8 & 71.1\rlap{$^*$} & 85.2 & 84.1 & 85.1 & 77.2 & 77.7\rlap{$^*$} & 89.5 & 83.4 & 85.8 & 82.9 \\
\quad SFT & 75.6 & 89.8 & 74.0 & 78.4 & 81.8 & 88.0\rlap{$^*$} & 94.9 & 93.5 & 95.2 & 91.0 & 73.7\rlap{$^*$} & 88.2 & 87.6 & 88.6 & 80.0 & 79.1\rlap{$^*$} & 91.0 & 85.0 & 87.4 & 84.3 \\
\quad SFT (1/4 data) + RL & 75.8 & \textbf{90.4} & 73.8 & 78.8 & 81.9 & 85.1\rlap{$^*$} & 96.4 & 94.0 & 97.1 & 90.0 & 74.1\rlap{$^*$} & 89.8 & 89.6 & 90.8 & 81.0 & 78.3\rlap{$^*$} & 92.2 & 85.8 & 88.9 & 84.3 \\
\quad SFT (1/2 data) + RL & \textbf{76.2} & 89.7 & 74.0 & 78.7 & \textbf{82.2} & 88.9 & 95.7 & 94.3 & 96.2 & 91.9 & 74.9\rlap{$^*$} & 90.2 & 90.5 & 91.9 & 81.6 & 80.0\rlap{$^*$} & 91.9 & 86.3 & 88.9 & 85.2 \\
Teacher & 75.9 & 89.3 & \textbf{74.1} & 78.6 & 82.0 & \textbf{89.1} & \textbf{97.2} & \textbf{95.2} & \textbf{97.3} & \textbf{92.7} & 73.7\rlap{$^*$} & 88.6 & 88.8 & 89.9 & 80.1 & 79.6\rlap{$^*$} & 91.7 & 86.0 & 88.6 & 84.9 \\
\midrule
\multicolumn{21}{l}{\textit{w/o reranker}} \\
\midrule
GRASP (SFT + RL) & 60.5 & \textbf{84.5} & \textbf{69.7} & \textbf{79.0} & 71.3 & \textbf{80.7} & 91.9 & 92.7 & 96.4 & \textbf{85.8} & \textbf{62.0} & \textbf{83.6} & \textbf{87.8} & \textbf{93.0} & \textbf{71.7} & \textbf{67.7} & \textbf{86.7} & \textbf{83.4} & \textbf{89.5} & \textbf{76.3} \\
\quad SFT (1/4 data) & 59.7 & 83.0 & 67.5 & 77.1 & 70.2 & 66.1\rlap{$^*$} & 82.5 & 85.5 & 90.2 & 73.9 & 40.3\rlap{$^*$} & 67.4 & 73.1 & 80.6 & 52.6 & 55.4\rlap{$^*$} & 77.6 & 75.4 & 82.6 & 65.6 \\
\quad SFT (1/2 data) & 59.7 & 83.7 & 68.7 & 78.2 & 70.5 & 77.1\rlap{$^*$} & 88.3 & 89.5 & 94.1 & 82.3 & 45.6\rlap{$^*$} & 68.8 & 77.4 & 85.1 & 56.1 & 60.8\rlap{$^*$} & 80.3 & 78.5 & 85.8 & 69.6 \\
\quad SFT & 60.2 & 83.9 & 69.3 & 78.4 & 70.8 & 79.4\rlap{$^*$} & 90.2 & 91.0 & 95.2 & 84.3 & 50.4\rlap{$^*$} & 75.9 & 81.8 & 88.6 & 62.0 & 63.3\rlap{$^*$} & 83.3 & 80.7 & 87.4 & 72.4 \\
\quad SFT (1/4 data) + RL & 60.2 & 83.6 & 68.9 & 78.8 & 70.8 & 70.5\rlap{$^*$} & 86.6 & 92.4 & 97.1 & 78.4 & 53.2\rlap{$^*$} & 80.0 & 84.8 & 90.8 & 65.0 & 61.3\rlap{$^*$} & 83.4 & 82.0 & 88.9 & 71.4 \\
\quad SFT (1/2 data) + RL & \textbf{60.9} & 84.2 & 69.2 & 78.7 & \textbf{71.4} & 80.1 & 91.6 & 92.2 & 96.2 & 85.4 & 58.5\rlap{$^*$} & 80.0 & 86.7 & 91.9 & 68.1 & 66.5\rlap{$^*$} & 85.3 & 82.7 & 88.9 & 75.0 \\
Teacher & 60.1 & 83.7 & 69.4 & 78.6 & 70.9 & 78.9\rlap{$^*$} & \textbf{93.5} & \textbf{93.4} & \textbf{97.3} & 85.1 & 53.5\rlap{$^*$} & 75.8 & 83.7 & 89.9 & 64.0 & 64.2\rlap{$^*$} & 84.3 & 82.2 & 88.6 & 73.3 \\
\bottomrule
\end{tabular}

    }
    \caption{Ablation on the graph branch training.
    SFT is trained on teacher-generated data, and 1/4 and 1/2 denote the fraction of the full SFT set ($N{=}500$) used.
    The base model without SFT fails to generate executable plans and is therefore omitted.
    Teacher uses the teacher model as the graph branch.
    \textbf{Bold} indicates the best result per metric per dataset within each setting.
    $^*$ marks H@1 results that differ significantly from GRASP in the same setting ($p<0.05$, McNemar's test, Stouffer's method for Average). A visualization is shown in Figure~\ref{fig:sft_scaling}.}
    \label{tab:ablation_graph_training}
\end{table*}

\begin{figure}[t]
    \centering
    \includegraphics[width=\columnwidth]{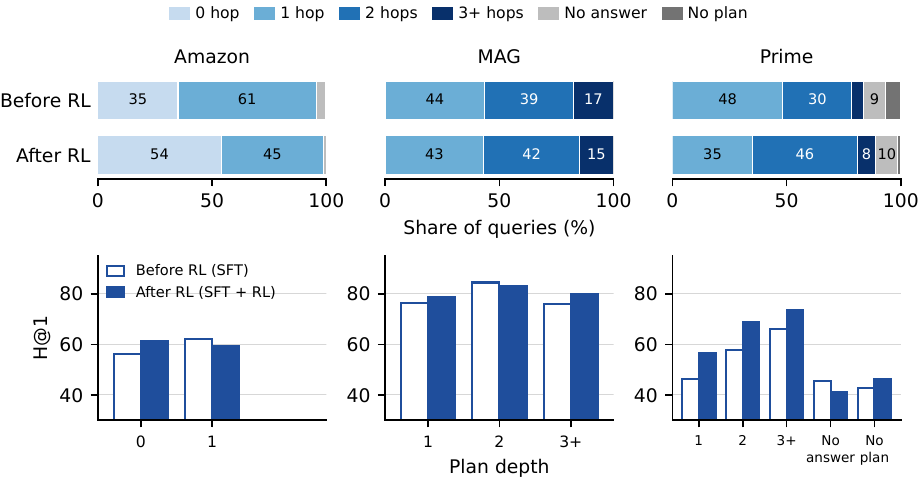}
    \caption{Plan depth analysis of the graph branch before RL (SFT) and after RL (SFT + RL), without the reranker.
    Top: share of queries by plan depth.
    No answer denotes plans that execute but return no answer node, and No plan denotes failed plan generation, in which case only the text branch is used.
    Bottom: H@1 for each plan depth, shown only for depths with at least 30 queries both before and after RL.}
    \label{fig:hop_analysis}
\end{figure}

\begin{table*}[t]
\centering
    \centering
    \resizebox{\ifdim\width>\textwidth \textwidth \else \width \fi}{!}{
        \begin{tabular}{l|cccc|cccc|cccc}
\toprule
\multirow{3}{*}{\textbf{Plan depth}} & \multicolumn{4}{c|}{\textbf{Amazon}} & \multicolumn{4}{c|}{\textbf{MAG}} & \multicolumn{4}{c}{\textbf{Prime}} \\
\cmidrule(lr){2-5} \cmidrule(lr){6-9} \cmidrule(lr){10-13}
& \multicolumn{2}{c}{Before RL} & \multicolumn{2}{c|}{After RL} & \multicolumn{2}{c}{Before RL} & \multicolumn{2}{c|}{After RL} & \multicolumn{2}{c}{Before RL} & \multicolumn{2}{c}{After RL} \\
\cmidrule(lr){2-3} \cmidrule(lr){4-5} \cmidrule(lr){6-7} \cmidrule(lr){8-9} \cmidrule(lr){10-11} \cmidrule(lr){12-13}
& \% & H@1 & \% & H@1 & \% & H@1 & \% & H@1 & \% & H@1 & \% & H@1 \\
\midrule
0 hop & 35.1 & 56.1 & 54.0 & 61.4 & -- & -- & -- & -- & -- & -- & -- & -- \\
1 hop & 60.6 & 62.2 & 44.8 & 59.5 & 43.6 & 76.4 & 43.1 & 78.8 & 48.3 & 46.2 & 35.1 & 56.7 \\
2 hops & -- & -- & -- & -- & 38.9 & 84.5 & 42.0 & 82.9 & 30.2 & 57.6 & 45.7 & 68.9 \\
3+ hops & -- & -- & -- & -- & 17.4 & 76.0 & 14.8 & 80.0 & 5.2 & 66.2 & 8.2 & 73.6 \\
No answer & 3.8 & 65.1 & 1.2 & 55.0 & -- & -- & -- & -- & 9.4 & 45.5 & 9.5 & 41.2 \\
No plan & 0.5 & 62.5 & -- & -- & 0.2 & 50.0 & 0.1 & 50.0 & 6.9 & 42.7 & 1.4 & 46.2 \\
\midrule
Total & 100.0 & 60.2 & 100.0 & 60.5 & 100.0 & 79.4 & 100.0 & 80.7 & 100.0 & 50.4 & 100.0 & 62.0 \\
\bottomrule
\end{tabular}
    }
    \caption{Plan depth distribution and H@1 of the graph branch before RL (SFT) and after RL (SFT + RL), without the reranker.
    \% denotes the share of queries, and H@1 is computed within each group.
    No answer denotes plans that execute but return no answer node, and No plan denotes failed plan generation, in which case only the text branch is used.
    The groups cover all 1,642 (Amazon), 2,665 (MAG), and 2,801 (Prime) test queries, so the totals match Table~\ref{tab:ablation_graph_training}.}
    \label{tab:hop_analysis}
\end{table*}

\begin{table*}[t]
\centering
    \centering
    \resizebox{\ifdim\width>\textwidth \textwidth \else \width \fi}{!}{
        \begin{tabular}{l|ccccc|ccccc|ccccc|ccccc}
\toprule
\multirow{2}{*}{\textbf{Variant}} & \multicolumn{5}{c|}{\textbf{Amazon}} & \multicolumn{5}{c|}{\textbf{MAG}} & \multicolumn{5}{c|}{\textbf{Prime}} & \multicolumn{5}{c}{\textbf{Average}} \\
\cmidrule(lr){2-6} \cmidrule(lr){7-11} \cmidrule(lr){12-16} \cmidrule(lr){17-21}
& \textbf{H@1} & \textbf{H@5} & \textbf{R@20} & \textbf{R@50} & \textbf{MRR} & \textbf{H@1} & \textbf{H@5} & \textbf{R@20} & \textbf{R@50} & \textbf{MRR} & \textbf{H@1} & \textbf{H@5} & \textbf{R@20} & \textbf{R@50} & \textbf{MRR} & \textbf{H@1} & \textbf{H@5} & \textbf{R@20} & \textbf{R@50} & \textbf{MRR} \\
\midrule
\multicolumn{21}{l}{\textit{w/ reranker ($K{=}50$)}} \\
\midrule
GRASP (full, mFAR) & \textbf{75.5} & \textbf{89.3} & \textbf{74.1} & \textbf{79.0} & \textbf{81.8} & \textbf{89.1} & \textbf{95.8} & \textbf{94.3} & \textbf{96.4} & \textbf{92.1} & \textbf{77.3} & \textbf{92.1} & \textbf{92.0} & \textbf{93.0} & \textbf{83.7} & \textbf{80.6} & \textbf{92.4} & \textbf{86.8} & \textbf{89.5} & \textbf{85.9} \\
GRASP (Qwen3) & 70.5\rlap{$^*$} & 84.4 & 67.0 & 70.8 & 76.7 & 88.2\rlap{$^*$} & 95.4 & 93.8 & 95.7 & 91.4 & 72.4\rlap{$^*$} & 85.2 & 84.0 & 85.3 & 78.1 & 77.0\rlap{$^*$} & 88.3 & 81.6 & 83.9 & 82.1 \\
\midrule
\multicolumn{21}{l}{\textit{w/o reranker}} \\
\midrule
GRASP (mFAR) & \textbf{60.5} & \textbf{84.5} & \textbf{69.7} & \textbf{79.0} & \textbf{71.3} & \textbf{80.7} & \textbf{91.9} & \textbf{92.7} & \textbf{96.4} & \textbf{85.8} & \textbf{62.0} & \textbf{83.6} & \textbf{87.8} & \textbf{93.0} & \textbf{71.7} & \textbf{67.7} & \textbf{86.7} & \textbf{83.4} & \textbf{89.5} & \textbf{76.3} \\
GRASP (Qwen3) & 46.1\rlap{$^*$} & 72.5 & 60.6 & 70.8 & 58.1 & 80.4 & \textbf{91.9} & 92.1 & 95.7 & 85.7 & 56.6\rlap{$^*$} & 75.5 & 79.8 & 85.3 & 65.0 & 61.0\rlap{$^*$} & 80.0 & 77.5 & 83.9 & 69.6 \\
\bottomrule
\end{tabular}
    }
    \caption{Ablation on the text branch.
    The text branch uses mFAR or Qwen3, as indicated in parentheses.
    \textbf{Bold} indicates the best result per metric per dataset within each setting.
    $^*$ marks H@1 results that differ significantly from the mFAR variant in the same setting ($p<0.05$, McNemar's test, Stouffer's method for Average).}
    \label{tab:ablation_text}
\end{table*}

\begin{figure*}[t]
    \centering
  \begin{subfigure}{0.32\linewidth}
    \includegraphics[width=\linewidth]{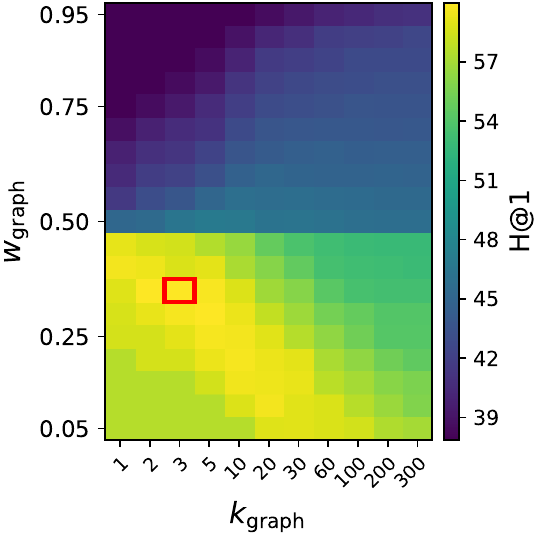}
    \caption{Amazon}
  \end{subfigure}
    \hfill
  \begin{subfigure}{0.32\linewidth}
    \includegraphics[width=\linewidth]{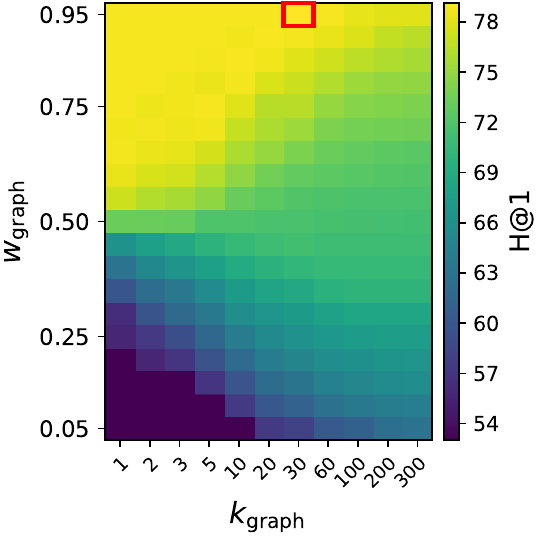}
    \caption{MAG}
  \end{subfigure}
    \hfill
  \begin{subfigure}{0.32\linewidth}
    \includegraphics[width=\linewidth]{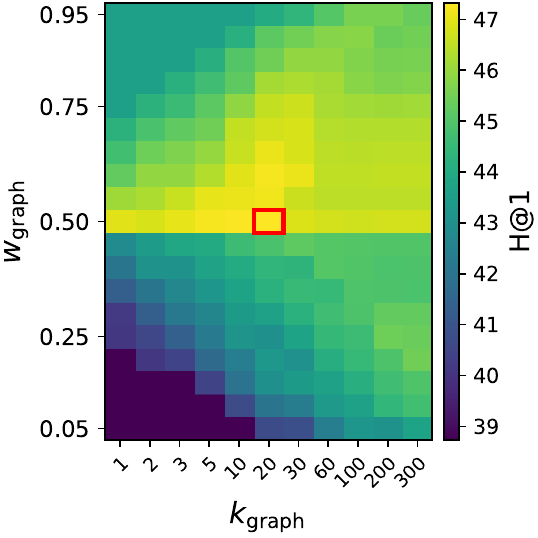}
    \caption{Prime}
  \end{subfigure}
    \caption{Validation H@1 over the RRF grid for fusing the SFT graph branch with the text branch. $w_\mathrm{graph}$ is the weight of the graph branch. The red box marks the selected optimal $(w, k)$ per dataset.}
    \label{fig:hyper_graph}
\end{figure*}

\begin{figure*}[t]
    \centering
  \begin{subfigure}{0.32\linewidth}
    \includegraphics[width=\linewidth]{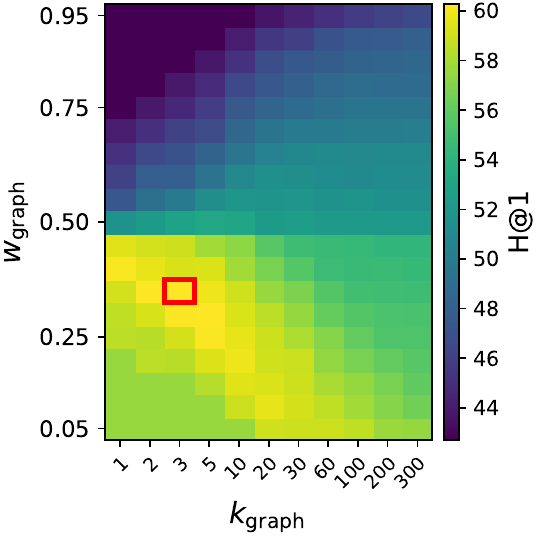}
    \caption{Amazon}
  \end{subfigure}
    \hfill
  \begin{subfigure}{0.32\linewidth}
    \includegraphics[width=\linewidth]{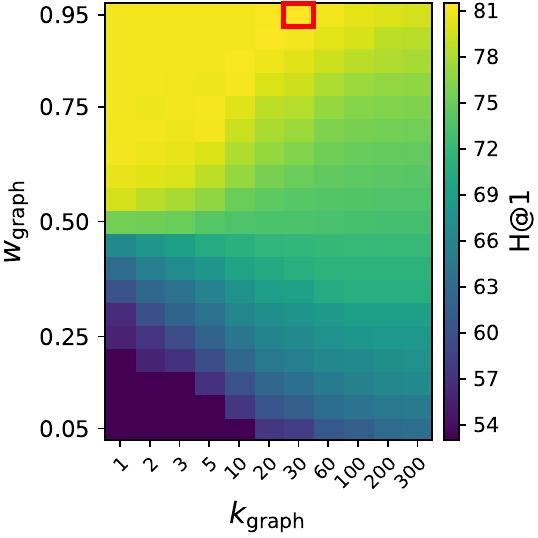}
    \caption{MAG}
  \end{subfigure}
    \hfill
  \begin{subfigure}{0.32\linewidth}
    \includegraphics[width=\linewidth]{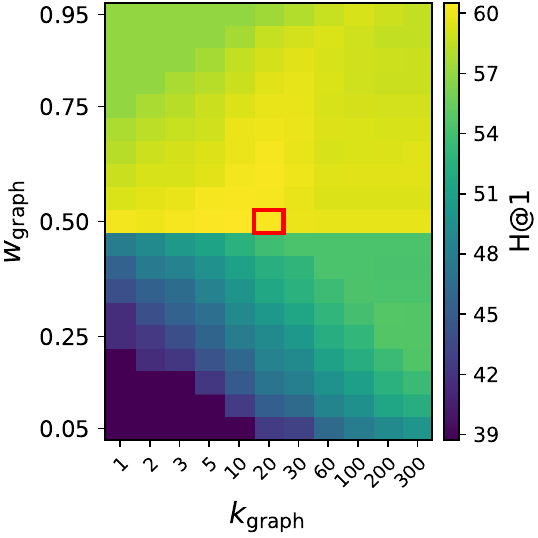}
    \caption{Prime}
  \end{subfigure}
    \caption{Validation H@1 over the RRF grid for fusing the SFT + RL graph branch with the text branch. The hyperparameters selected on the SFT models (Figure~\ref{fig:hyper_graph}) are kept, which remain optimal or within 0.3 H@1 of the optimums after RL training.}
    \label{fig:hyper_graph_trained}
\end{figure*}

\begin{figure*}[t]
    \centering
  \begin{subfigure}{0.32\linewidth}
    \includegraphics[width=\linewidth]{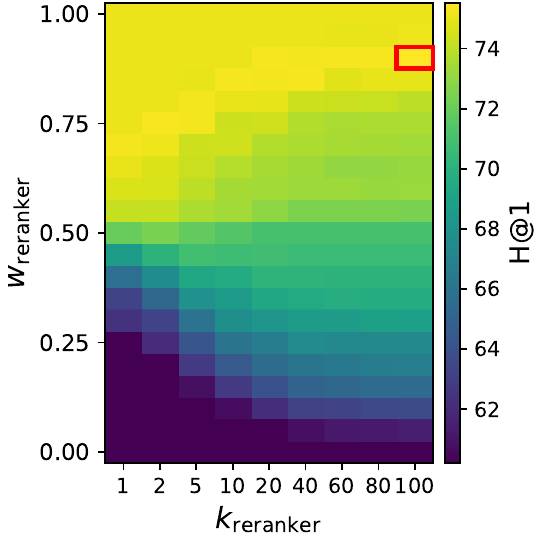}
    \caption{Amazon}
  \end{subfigure}
    \hfill
  \begin{subfigure}{0.32\linewidth}
    \includegraphics[width=\linewidth]{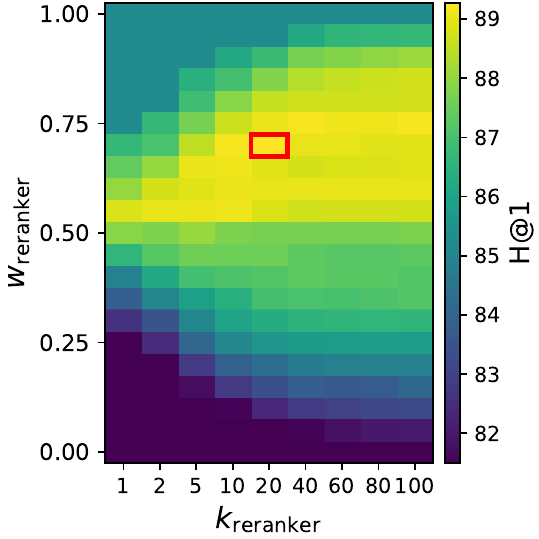}
    \caption{MAG}
  \end{subfigure}
    \hfill
  \begin{subfigure}{0.32\linewidth}
    \includegraphics[width=\linewidth]{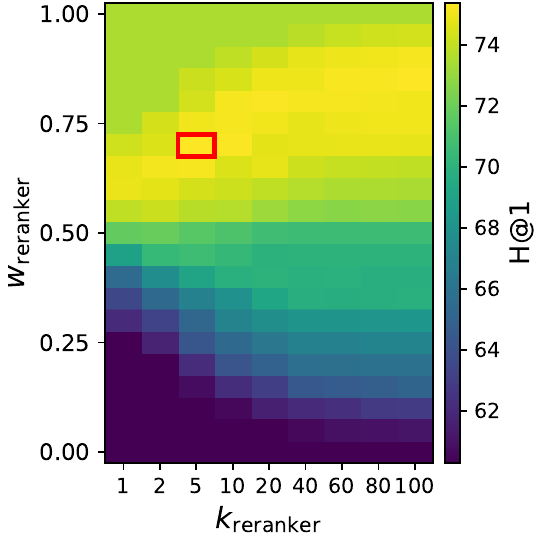}
    \caption{Prime}
  \end{subfigure}
    \caption{Validation H@1 over the RRF grid for fusing the reranker with the fused graph-text branch. $w_\mathrm{reranker}$ is the weight for the reranker. The red box marks the selected optimal $(w, k)$ per dataset.}
    \label{fig:hyper_reranker}
\end{figure*}

\section{Baselines}
\label{app:exp_baselines}

We compare against widely used text-based and graph-based baselines. Text-based: BM25~\cite{bm25}, (multi-)ada-002~\cite{ada002}, DPR with a RoBERTa encoder~\cite{dpr}, Qwen3-Embedding-4B~\cite[Qwen3]{qwen3}, KAR~\cite{kar}, and mFAR~\cite{mfar}. Graph-based: QAGNN~\cite{qagnn}, ToG~\cite{tog}, AvaTaR~\cite{avatar}, HybGRAG~\cite{hybgrag}, MoR~\cite{mor}, GraphRAFT~\cite{graphraft}, and AF-Retriever~\cite{afretriever}.

We take the results of BM25, ada-002, multi-ada-002, DPR, and QAGNN from STaRK, those of ToG from MoR, and those of AvaTaR, KAR, mFAR, HybGRAG, MoR, GraphRAFT, and AF-Retriever from their own papers. We evaluate Qwen3 using the official checkpoint. We re-run mFAR and report the results as mFAR$^*$ alongside the originally reported numbers. Unless otherwise stated, mFAR in the text refers to mFAR$^*$.

\section{Additional Implementation Details}
\label{app:impl}

\paragraph{Models.}
The entity extractor and the planner are both Qwen3-1.7B, used with thinking disabled.
The extractor is the base model, prompted with four demonstrations, and the planner is fine-tuned as described in Section~\ref{sec:method_stage2}.
The teacher that writes the supervised plans is Qwen3.6-27B.
In the graph branch, anchors, relevance texts, and nodes are embedded with Qwen3-Embedding-4B on all three datasets (last-token pooling, no instruction prefix).
The text branch is mFAR \citep{mfar} trained on each dataset's training split.
The reranker is Qwen3-Reranker-0.6B, and the neighbors listed in its documents are ranked by Qwen3-Embedding-4B cosine similarity to the query, without an instruction prefix.

\paragraph{Knowledge base and anchor linking.}
The SKB is hosted on Neo4j \citep{neo4j}.
Every node carries a name embedding and a document embedding, pre-computed at indexing time and stored in one name index and one document index per node type.
Names are embedded with at most 64 tokens and documents with at most 512 tokens.
The top-$k$ matches whose cosine similarity is at least $\tau$ times the top-1's similarity become candidate bindings for that anchor variable, each with its cosine score.
We use $(k, \tau) = (5, 0.95)$ for name-mode and $(10, 0.90)$ for doc-mode, identical across the three datasets.
Entity linking for the entity table uses the same name indexes and keeps the top 3 nodes per node type.
Dense rescoring computes exact cosine similarities for up to 20{,}000 candidates (beyond that, through a nearest-neighbor search with the same depth), and the graph ranking keeps the top 100.

\paragraph{Planner training.}
The supervised data are the first 500 training queries, in a fixed random order, whose teacher plan is valid, places a correct answer first in its executed graph ranking, and whose query has at most 20 correct answers.
Supervised fine-tuning, used both for the warm start and for RFT, trains LoRA adapters ($r{=}64$, $\alpha{=}128$, all linear layers) for 3 epochs with a learning rate of $10^{-4}$, cosine decay, 3\% warmup, and an effective batch of 32.
In each round, eight plans per training query are sampled at temperature 1.0.
RFT keeps each query's highest-reward plan if its reward is positive, breaking ties by fewer anchors and then fewer candidates.
GRPO starts from the merged RFT model and trains new LoRA adapters ($r{=}32$, $\alpha{=}64$, all linear layers) with eight completions per prompt at temperature 1.0, 16 prompts per step, a learning rate of $2{\times}10^{-5}$, a KL coefficient of $0.01$, rewards that are not normalized by the group's standard deviation, and at most 512 completion tokens.
A group whose eight rewards are all equal is resampled up to three times.
Training runs for up to 200 steps, a checkpoint is saved every 50 steps, and the one with the highest validation fused Hit@1 is kept.
Training queries with more than 20 correct answers are excluded from RFT and GRPO.
In the reward table $T$, a group with fewer than 50 validation queries is shrunk toward the average of its gold-rank group over all candidate counts, with weight $\min(1, n/50)$ on its own average.
At inference, the planner decodes greedily with at most 700 output tokens.
A failed attempt is followed by up to two samples at temperature 0.7.

\paragraph{Reranker training.}
LoRA adapters with rank $r{=}32$, scaling $\alpha{=}64$, and dropout $0.05$ are attached to all linear layers.
Training runs for 3 epochs with cosine learning rate decay from $3{\times}10^{-5}$, 5\% warmup, AdamW with $(\beta_1, \beta_2) = (0.9, 0.95)$, weight decay $0.1$, gradient clipping at $1.0$, and a global batch of 16 queries.
Each text field of a document is capped at 512 tokens, and the full input (instruction, query, and document) at 2048 tokens.
Inputs over this budget first drop neighbors from the last relation line backwards and then shorten the text fields.
A checkpoint is saved every 0.5 epoch, and together with the base model, these checkpoints are the candidates for the validation selection in Section~\ref{sec:method_stage3}.

\paragraph{Hardware.}
Experiments are conducted on a Linux server with a 16-core Intel Xeon Gold 6242R CPU, 256 GB RAM, and two NVIDIA H100 GPUs (80 GB VRAM each). Training the full pipeline on the largest dataset (MAG) takes about 20 GPU hours.
All results are from a single run.

\section{Why Does RL Not Help on Amazon?}
\label{app:nothelpamazon}
On Amazon, the planner never produces plans deeper than one hop, either before or after RL, consistent with the limited relational structure of its SKB.
RL further shifts plans toward zero-hop retrieval, raising their share from 35.1\% to 54.0\%.
A zero-hop plan skips the traversal and selects its candidates solely by document similarity to the relevance text, so after RL the graph branch acts as a dense text retriever for more than half of Amazon queries.
This improves zero-hop accuracy (56.1 to 61.4) but lowers one-hop accuracy (62.2 to 59.5), leaving the overall H@1 nearly unchanged (60.2 vs.\ 60.5).
Such retrieval is largely redundant with the text branch.
The text branch alone already matches the full system on Amazon (Table~\ref{tab:ablation_branch}), and removing the relevance text leaves Amazon unchanged (Table~\ref{tab:ablation_graph_design}), since zero-hop plans then simply retrieve with the full query, which is equally redundant with the text branch.
Hence, improvements to the graph branch have little room to affect the final ranking on Amazon.

\section{Hyperparameter Sensitivity Analysis}
\label{app:exp_sensitivity}

The main hyperparameters we tune for GRASP on the validation set are the $(w, k)$ of the two fusions: one between the SFT graph branch and the text branch (Figure~\ref{fig:hyper_graph}), and the other between the reranker and the fused graph--text branch (Figure~\ref{fig:hyper_reranker}).
In general, performance remains stable around the chosen settings as long as the preferred branch is unchanged, suggesting that GRASP is not sensitive to these hyperparameters.
We examine these two fusions in more detail below.

For the former fusion, different branches clearly excel on different datasets, i.e., the text branch on Amazon and the graph branch on MAG and Prime, which makes their fusion necessary to generalize across domains.
Performance also changes sharply at $w_{\text{graph}}{=}0.5$, where the preferred branch switches, reflecting a substantial difference between the two branches: fusing them is not a simple aggregation of homogeneous methods, but a combination of orthogonal strengths.
We then check whether, after RL training, the SFT graph branch still fuses well with the text branch under the same fusion setting (Figure~\ref{fig:hyper_graph_trained}).
The hyperparameters selected before RL remain optimal or near-optimal, showing that, with a properly designed reward, RL training indeed strengthens the specific fusion the hyperparameters define.
Meanwhile, RL largely preserves the relative performance distribution over hyperparameters, indicating that it does not cause overfitting to the chosen fusion setting but improves the general capability of the graph branch.

For the latter fusion, the reranker carries more weight on all three datasets, confirming that reranking itself improves performance.
The performance drop across $w_{\text{reranker}}{=}0.5$ further reveals its superiority over the fused ranking.
Nevertheless, the chosen weight is never 1, implying that the fused branch is still valuable for deriving candidates directly from the query. This is most evident on MAG, where using the reranker alone is clearly suboptimal.

\section{Prompts}

\begin{figure*}[p]
\begin{Verbatim}
You extract entity mentions from {questions} so they can be linked to nodes of {kg} by name similarity.

Node types:
{node_types}

For each mention give:
- text: the mention written the way a graph node name would be written. Copy the question's wording when it already is a name. Rewrite only when needed: split coordinated phrases into separate mentions and restore a name that the question paraphrases.{text_rules}
- types: the 1 or 2 most likely node types.
- role: "anchor" if it names a specific entity the answer is related to; "target" if it describes the answer itself (what is asked for); "attribute" if it is a property or constraint that is not itself a graph node.{role_rules}
- parent: if the mention is nested inside a longer mention in your list, the index (0-based) of that longer mention, otherwise -1.
List only what the question itself mentions.

Output JSON only: {"mentions": [...]}
\end{Verbatim}
\vspace{2pt}
\begin{Verbatim}
Question: {question}
\end{Verbatim}
\vspace{2pt}
\begin{Verbatim}
{questions}  = a product search query
{kg}         = a product knowledge graph (Amazon)
{node_types} =
- product: a specific product, named by its title or model (e.g. "Fitbit Flex", "GSI Outdoors Glacier Stainless Steel Cookset")
- brand: a brand, manufacturer or licensed team (e.g. "Nike", "New Era", "Atlanta Braves")
- category: a kind of product (e.g. "broadheads", "boxing gloves", "dry bags")
- color: a color or color pattern (e.g. "maroon", "black/white", "grey camo")
{text_rules} = For example, "grey and red shades" -> "grey", "red". Keep brands, colors and product names out of the target: a brand written before the product type is its own anchor ("a RUIMX waterproof LED flashlight" -> brand "RUIMX", target "LED flashlight").
{role_rules} = An anchor names a specific brand, color, category or product the answer is related to (made by it, in that color, bought or used together with it). A target is the kind of product asked for. An attribute is a concrete requirement that is not itself a graph node (e.g. "waterproof", "under $50", "for women"). Skip vague words such as "good", "high-quality", "recommended". List both a full product name and the brand inside it.
\end{Verbatim}
\vspace{2pt}
\begin{Verbatim}
User:      Question: What are some broadheads that are frequently purchased with the Dead Ringer White Tail Freaks Super-Freak 100 Grain Broadhead, Stainless?
Assistant: {"mentions": [{"text": "broadheads", "types": ["product", "category"], "role": "target", "parent": -1}, {"text": "Dead Ringer White Tail Freaks Super-Freak 100 Grain Broadhead, Stainless", "types": ["product"], "role": "anchor", "parent": -1}, {"text": "Dead Ringer", "types": ["brand"], "role": "anchor", "parent": 1}]}
\end{Verbatim}
\vspace{2pt}
\begin{Verbatim}
Question: What are some valuable and durable discus options for track and field similar to the Martin Sports Abs Plastic Discus, 1 kg/2.2 lbs. I previously purchased?
\end{Verbatim}
\vspace{2pt}
\begin{Verbatim}
{"mentions": [{"text": "discus", "types": ["product", "category"], "role": "target", "parent": -1}, {"text": "Martin Sports Abs Plastic Discus, 1 kg/2.2 lbs.", "types": ["product"], "role": "anchor", "parent": -1}, {"text": "Martin Sports", "types": ["brand"], "role": "anchor", "parent": 1}]}
\end{Verbatim}
\caption{Prompt for extracting entity mentions from a question. The system prompt is shared by all datasets: \texttt{\{questions\}}, \texttt{\{kg\}} and \texttt{\{node\_types\}} describe each dataset, and \texttt{\{text\_rules\}} and \texttt{\{role\_rules\}} add dataset-specific guidance to the shared rules. Four demonstrations from the training set precede the question as user and assistant turns. The model (Qwen3-1.7B) decodes greedily, and its output is constrained to the JSON schema. The example boxes show the Amazon slot values, one of the four Amazon demonstrations, and the output for the question of Figure~\ref{fig:plan-prompt}. After each mention is linked to graph nodes, the three mentions become entries E1--E3 of the entity table in Figure~\ref{fig:plan-prompt}.}
\label{fig:extract-prompt}
\end{figure*}

\begin{figure*}[p]
\begin{Verbatim}
You convert a question over {kg} into a graph-traversal plan, emitted as ONE JSON object.

## Node labels
{labels}

## Relations (all undirected)
{relations}

## Plan
- anchors: entities in the question that map to graph nodes: {"var","text","label","match_mode"}; match_mode is "name" for names and titles, "doc" for descriptive phrases.
- hops: traversals {"from","rel","to_var","to_label"}; "from" must be an already introduced variable. Hops may be [] only when there is exactly one anchor, its var is target.var and target.labels is [its label]; the answer is then retrieved by its description.
- target: the answer node {"var","labels","relevance_text"}; relevance_text is the part of the question describing the answer that the hops do not capture ("" if none).
All anchors must be satisfied jointly, so use only anchors the question really constrains.
Output only the JSON object with keys "anchors", "hops", "target".
\end{Verbatim}
\vspace{2pt}
\begin{Verbatim}
Question: {question}

Entities found in the question, each with the graph nodes it links to (similarity, node degree). Anchor text may copy a mention or a node name.
{entity_table}
\end{Verbatim}
\vspace{2pt}
\begin{Verbatim}
{kg}        = a product knowledge graph (Amazon)
{labels}    = brand, category, color, product
{relations} =
- brand -HAS_BRAND- product
- category -HAS_CATEGORY- product
- color -HAS_COLOR- product
- product -{ALSO_BUY|ALSO_VIEW}- product
\end{Verbatim}
\vspace{2pt}
\begin{Verbatim}
Question: What are some valuable and durable discus options for track and field similar to the Martin Sports Abs Plastic Discus, 1 kg/2.2 lbs. I previously purchased?

Entities found in the question, each with the graph nodes it links to (similarity, node degree). Anchor text may copy a mention or a node name.
E1 "discus" (target) -> product: Discus 1.6 Kilo (0.91, 4), Supra Black Discus High School Level 1.6 Kilo Discus (0.90, 5), Competitive Rubber 1.6 Kilo Discus (0.90, 5) | category: discuses (0.94, 109), discs (0.88, 1726), disc sports (0.88, 2954)
E2 "Martin Sports Abs Plastic Discus, 1 kg/2.2 lbs." (anchor) -> product: Martin Sports Abs Plastic Discus, 1 kg/2.2 lbs. (1.00, 36), Martin Sports Abs Plastic Discus, 1.6 kg/3.52 lbs. (1.00, 22), AMBER Athletic Gear Ultimate Discus, 1 Kg (0.96, 4)
E3 "Martin Sports" (anchor, inside E2) -> brand: Martin Sports (1.00, 64), Scott Sports (0.96, 58), ERS Sport (0.96, 1)
\end{Verbatim}
\vspace{2pt}
\begin{Verbatim}
{"anchors": [{"var": "A1", "text": "Martin Sports Abs Plastic Discus, 1 kg/2.2 lbs", "label": "product", "match_mode": "name"}], "hops": [{"from": "A1", "rel": "ALSO_VIEW", "to_var": "T", "to_label": "product"}], "target": {"var": "T", "labels": ["product"], "relevance_text": "valuable and durable discus options for track and field"}}
\end{Verbatim}
\caption{System and user prompts for converting a question into a graph-traversal plan. In the system prompt, \texttt{\{kg\}}, \texttt{\{labels\}} and \texttt{\{relations\}} are filled in with the schema of each knowledge graph. In the user prompt, \texttt{\{question\}} is the input question and \texttt{\{entity\_table\}} lists the entities found in it together with their candidate graph nodes. The three example boxes show, for a question from the Amazon dataset, the filled-in schema, the complete user prompt, and the plan generated by the model.}
\label{fig:plan-prompt}
\end{figure*}

\begin{figure*}[p]
\begin{Verbatim}
Judge whether the Document meets the requirements based on the Query and the Instruct provided. Note that the answer can only be "yes" or "no".
\end{Verbatim}
\vspace{2pt}
\begin{Verbatim}
<Instruct>: Given a question about entities in {domain} knowledge graph, judge whether the candidate node is a correct answer. Consider the node's attributes, its relations to other nodes, and how well it matches the question's requirements.
<Query>: {question}
<Document>: {candidate}
\end{Verbatim}
\vspace{2pt}
\begin{Verbatim}
{domain}    = a product
{candidate} =
[Type] <node label>
[name] <node name>
[details] <text fields of the node>
[<relation>] <neighbor>; <neighbor>; ...     (one line per relation; up to 10 neighbors, most similar to the question first)
\end{Verbatim}
\vspace{2pt}
\begin{Verbatim}
<Instruct>: Given a question about entities in a product knowledge graph, judge whether the candidate node is a correct answer. Consider the node's attributes, its relations to other nodes, and how well it matches the question's requirements.
<Query>: What are some valuable and durable discus options for track and field similar to the Martin Sports Abs Plastic Discus, 1 kg/2.2 lbs. I previously purchased?
<Document>: [Type] product
[name] Champro Rubber Discus
[details] product: Champro Rubber Discus. brand: CHAMPRO. features: molded rubbber discus | official size and weight | great practice discus for indoors and outdoors. reviews: it's a rubber disc. | The first one that I received had a huge chunk ... | Great Budget Training Discus | Like his disc at school | durable
[ALSO_BUY] Offical Black Rubber Discus - 1K; Champion Sports Shot/Discus Carrier with Shoulder Strap; Port a Pit Shot And Discus Carrier; Scholastic Iron Shot Put - 8LBS; Competition Shot Put, 4k; Scholastic Iron Shot Put - 6LBS; Competition Shot Put, 12 lbs; Cintz Anodized Aluminium Relay Batons, set of 8, Assorted colors
[ALSO_VIEW] Martin Sports Abs Plastic Discus, 1 kg/2.2 lbs.
[HAS_BRAND] CHAMPRO
[HAS_CATEGORY] throwing equipment; discuses; other team sports
\end{Verbatim}
\vspace{2pt}
\begin{Verbatim}
Champro Rubber Discus (correct answer):
    logit("yes") - logit("no") = 8.09
Martin Sports Abs Plastic Discus, 1 kg/2.2 lbs. (already purchased, not an answer):
    logit("yes") - logit("no") = 3.45
\end{Verbatim}
\caption{Prompt of the reranker, Qwen3-Reranker-0.6B fine-tuned with LoRA. The system prompt and the \texttt{<Instruct>}\,/\,\texttt{<Query>}\,/\,\texttt{<Document>} layout are those of Qwen3-Reranker. We set only the instruction, which differs across datasets only in \texttt{\{domain\}} (a product, an academic or a biomedical). Each of the top 50 fused candidates is written as its type, name and text fields, followed by one line per relation type. Each line lists up to ten neighbors, ordered by their embedding similarity to the question. The score of a candidate is $\mathrm{logit}(\texttt{yes}) - \mathrm{logit}(\texttt{no})$ for the next token, and candidates are ranked by this score. The example boxes show the Amazon instruction, the candidate format and the complete input for one candidate of the question in Figure~\ref{fig:plan-prompt}. The output box shows the scores of the Amazon reranker for a correct answer and for the product the user already purchased, which is not an answer.}
\label{fig:rerank-prompt}
\end{figure*}

\end{document}